\documentclass[11pt,prd,onecolumn,showpacs,amsmath,amssymb,aps,floats,floatfix]{revtex4-1}

\usepackage[inline]{enumitem}
\usepackage[multidot]{grffile}  
\usepackage{dcolumn}
\usepackage{bm}
\usepackage{amsmath}
\usepackage{amsfonts}
\usepackage{amssymb}
\usepackage{color}
\usepackage{latexsym}
\usepackage{slashed} 
\usepackage{pstricks}
\usepackage{indentfirst}
\usepackage{mathrsfs}
\usepackage{multirow}
\usepackage{epsfig,psfrag}
\usepackage{subfigure}
\usepackage{mathtools}
\usepackage{setspace} 
\usepackage[utf8]{inputenc} 
\usepackage[scientific-notation=true]{siunitx} 

\graphicspath{{figs/}}

\def\pT{p_\mathrm{T}} 
\def\missET{\slashed E_\mathrm{T}} 

\renewcommand{\vec}[1]{\mathbf{#1}}

\newcommand{\SUtwoL}{\mathrm{SU}(2)_\mathrm{L}}

\makeatother

\allowdisplaybreaks 

\begin{document}

\title{Exploring triplet-quadruplet fermionic dark matter at the LHC and future colliders}
\author{Jin-Wei Wang$^{1,2}$}
\email{wangjinwei@ihep.ac.cn}
\author{Xiao-Jun Bi$^1$}
\email{bixj@ihep.ac.cn}
\author{Qian-Fei Xiang$^{1,2}$}
\email{xiangqf@ihep.ac.cn}
\author{Peng-Fei Yin$^1$}
\email{yinpf@ihep.ac.cn}
\author{Zhao-Huan Yu$^{3,4}$}
\email{zhao-huan.yu@unimelb.edu.au}
\affiliation{$^1$Key Laboratory of Particle Astrophysics,
Institute of High Energy Physics, Chinese Academy of Sciences,
Beijing 100049, China}
\affiliation{$^2$School of Physical Sciences,
University of Chinese Academy of Sciences,
Beijing 100049, China}
\affiliation{$^3$ARC Centre of Excellence for Particle Physics at the Terascale,
School of Physics, The~University of Melbourne, Victoria 3010, Australia}
\affiliation{$^4$School of Physics, Sun Yat-Sen University, Guangzhou 510275, China}

\begin{abstract}
We study the signatures of the triplet-quadruplet dark matter model at the LHC and future colliders, including the 100~TeV Super Proton-Proton Collider and the 240~GeV Circular Electron Positron Collider.
The dark sector in this model contains one fermionic electroweak triplet and two fermionic quadruplets, which have two kinds of Yukawa couplings to the Higgs doublet.
Electroweak production signals of the dark sector fermions in the $\text{monojet}+\missET$, disappearing track, and $\text{multilepton}+\missET$ channels at the LHC and the Super Proton-Proton Collider are investigated.
Moreover, we study the loop effects of this model on the Circular Electron Positron Collider precision measurements of $e^+e^-\to Zh$ and $h\to\gamma\gamma$.
We find that most of the parameter regions allowed by the observed dark matter relic density will be well explored by such direct and indirect searches at future colliders.
\end{abstract}

\maketitle
\tableofcontents
\clearpage

\section{Introduction}
With the discovery of the Higgs boson at the Large Hadron Collider (LHC) \cite{Aad:2012tfa,Chatrchyan:2012xdj}, the last missing piece of the standard model (SM) has been found. However, solid astrophysical and cosmological observations reveal the existence of dark matter (DM),
which opens a door for exploring new physics beyond the standard model (BSM).
Among various DM candidates proposed, weakly interacting massive particles (WIMPs) are very compelling, because they can naturally explain the DM relic density with particle masses of $\mathcal{O}(\si{GeV}) - \mathcal{O}(\si{TeV})$ \cite{Bertone:2004pz,Feng:2010gw,Young:2016ala,Arcadi:2017kky}.

Many new physics models motivated by deep theoretical problems, \textit{e.g.}, supersymmetry (SUSY) models~\cite{Jungman:1995df}, naturally provide viable WIMP candidates.
Nonetheless, in spite of other motivations, WIMP models can be easily constructed by introducing a dark sector with electroweak (EW) $\SUtwoL$ multiplets, whose neutral components provide a potential DM candidate.
A dark sector containing one nontrivial $\SUtwoL$ multiplet is considered a minimal extension, leading to the so-called minimal dark matter models~\cite{Cirelli:2005uq,Cirelli:2009uv,Hambye:2009pw,Cai:2012kt,Ostdiek:2015aga,Cai:2015kpa,DelNobile:2015bqo}. Furthermore, introducing more than one $\SUtwoL$ multiplet in the dark sector gives rise to a much richer phenomenology~\cite{Mahbubani:2005pt,DEramo:2007anh,Enberg:2007rp,Cohen:2011ec,Fischer:2013hwa,Cheung:2013dua,Dedes:2014hga,Fedderke:2015txa,Calibbi:2015nha,Freitas:2015hsa,Yaguna:2015mva,Tait:2016qbg,Horiuchi:2016tqw,Banerjee:2016hsk,Cai:2016sjz,Abe:2017glm,Lu:2016dbc,Cai:2017wdu,Maru:2017otg,Liu:2017gfg,Egana-Ugrinovic:2017jib,Xiang:2017yfs}.

Among EW gauge eigenstates in SUSY models, bino is an $\SUtwoL$ singlet, Higgsinos belong to a doublet, winos form a triplet.
Thus, a dark sector with singlet and doublet fermions~\cite{Mahbubani:2005pt,DEramo:2007anh,Enberg:2007rp,Cohen:2011ec,Cheung:2013dua,Calibbi:2015nha,Horiuchi:2016tqw,Banerjee:2016hsk,Cai:2016sjz,Abe:2017glm,Xiang:2017yfs}, or with doublet and triplet fermions~\cite{Dedes:2014hga,Cai:2016sjz,Xiang:2017yfs,Voigt:2017vfz}, is analogous to the EW sector of SUSY models in special limits, which has been well studied in the literature for decades.
A more complicated case with triplet and quadruplet fermions~\cite{Tait:2016qbg,Cai:2016sjz}, however, cannot be regarded as a limit of the SUSY electroweak sector.
Thus, we may expect that the phenomenology could be quite different.

In such a triplet-quadruplet dark matter (TQDM) model, the dark sector involves one Weyl triplet with $Y=0$ and two Weyl quadruplets with $Y=\pm 1/2$.
After the electroweak symmetry breaking (EWSB), these multiplets mix with each other; the mass eigenstates include
three neutral Majorana fermions $\chi_{1,2,3}^0$, three singly charged fermions $\chi_{1,2,3}^\pm$, and one doubly charged fermion $\chi^{\pm\pm}$.
By imposing a discrete $Z_2$ symmetry, the lightest neutral fermion $\chi_1^0$ is stable, serving as a DM candidate. 
In this work, we investigate signatures of the TQDM model at the LHC and future colliders.

Through their EW gauge interactions, these dark sector fermions can be directly produced in $pp$ collisions,
leading to unique signatures at the LHC, as well as at future $\sim 100~\si{TeV}$ $pp$ colliders, such as the Super Proton-Proton Collider (SPPC)~\cite{CEPC-SPPCStudyGroup:2015csa,CEPC-SPPCStudyGroup:2015esa} and the Future Circular Collider with hadron collisions (FCC-hh)~\cite{Mangano:2017tke}.
All dark sector fermions would sequentially decay into the DM candidate $\chi_1^0$, which escapes detection and leads to a significant amount of missing transverse energy ($\missET$).
Therefore, the $\text{monojet}+\missET$ channel~\cite{Beltran:2010ww,Fox:2011pm,Aaboud:2016tnv,ATLAS:2017dnw}, where the final states are a large $\missET$ associated with $\geq 1$ hard jet from initial state radiation (ISR), should be very efficient for tagging such signal events. 

Besides, if $\chi_1^0$ consists of the pure triplet component or the  pure quadruplet components, the mass splitting between $\chi_1^\pm$ and $\chi_1^0$ would be $\sim \alpha_2 m_W \sin^2(\theta_\mathrm{W}/2) = 167~\si{MeV}$, only induced by loop effects~\cite{Cirelli:2005uq,Cirelli:2009uv,Tait:2016qbg}.
Consequently, the charged fermion $\chi_1^\pm$ has a macroscopic lifetime, and can travel a short distance in the inner detector before decaying to $\chi^0_1$ and a very soft, unlikely detected $\pi^\pm$ meson.
This causes a disappearing track signature that has been well studied at the LHC~\cite{Low:2014cba,ATLAS:2017bna,Aad:2013yna,Cirelli:2014dsa,Fukuda:2017jmk,Ostdiek:2015aga,Mahbubani:2017gjh}.
Furthermore, if the mass splittings between $\chi_1^0$ and other dark sector fermions are close to or larger than $m_W$ and $m_Z$, the $\text{multilepton}+\missET$ final states~\cite{ATLAS:2017uun,CMS:2017fdz} could be utilized to probe the TQDM model.

Several projects of high energy $e^+e^-$ colliders have been proposed, including the Circular Electron Positron Collider (CEPC)~\cite{CEPC-SPPCStudyGroup:2015csa,CEPC-SPPCStudyGroup:2015esa}, the International Linear Collider (ILC)~\cite{Baer:2013cma},
and the Future Circular Collider with $e^+e^-$ collisions (FCC-ee)~\cite{Gomez-Ceballos:2013zzn}.
At these $e^+e^-$ colliders, a lot of Higgs and EW measurements with unprecedentedly high precisions, providing excellent indirect approaches to BSM electroweak multiplets.
The sensitivity to the TQDM model via precision measurements of EW oblique parameters has been studied in Ref.~\cite{Cai:2016sjz}.
In this paper, we investigate the impact of the TQDM model on the Higgs physics at the CEPC, including the loop effects on the $e^+e^- \to Zh$ production~\cite{McCullough:2013rea,Shen:2015pha,Xiang:2017yfs} and the 
$h\to \gamma\gamma$ decay.

This paper is outlined as follows. In Sec.~\ref{sec:model} we give a brief description of the TQDM model and analyze its mass spectrum.
In Sec.~\ref{sec:LHC} we investigate current constraints from the $\text{monojet}+\missET$, disappearing track, and $\text{multilepton}+\missET$ channels at the 13~TeV LHC, and further study
the prospects at the 100~TeV SPPC based on Monte Carlo simulation.
In Sec.~\ref{sec:CEPC}, we calculate the one-loop corrections to the $e^+e^- \to Zh$ production cross section and $h\to \gamma\gamma$ partial width and estimate the CEPC prospects.
Conclusions and further discussions are given in Sec.~\ref{sec:conslusion}.

\section{Triplet-Quadruplet Dark Matter Model}
\label{sec:model}
\subsection{Model details}
In the TQDM model \cite{Tait:2016qbg,Cai:2016sjz}, the dark sector involves one colorless left-handed Weyl triplet $T$ and two colorless left-handed Weyl quadruplets $Q_1$ and $Q_2$ obeying the following $SU(2)_\text{L}\times U(1)_\text{Y}$ gauge transformations:
 \begin{equation}
   T = \left( \begin{array}{c}
      {T^ + }\\
      {T^0}\\
     -{T^-}
   \end{array} \right)
   \in (\bm{3},0)
   ,\hspace{0.5cm}
   Q_1 = \left( \begin{array}{c}
      {Q_1^ + }\\
      {Q_1^0}\\
      {Q_1^-}\\
      {Q_1^{--}}
   \end{array} \right)
   \in (\bm{4},-\frac 12)
   ,\hspace{0.5cm}
   Q_2 = \left( \begin{array}{c}
     {Q_2^{++}}\\
      {Q_2^+}\\
      {Q_2^0}\\
      {Q_2^-}
   \end{array} \right)
   \in (\bm{4}, \frac 12).
   \label{1}
 \end{equation}
The hypercharge signs of the two quadruplets are opposite, which makes sure that the TQDM model is anomaly free. Gauge-invariant Lagrangians for the triplet and quadruplets are given by
 \begin{equation}
   \mathcal{L}_\text{T}=i{T^\dag} \bar{\sigma}^\mu {D_\mu }T - ({m_T} a_{ij} T^i T^j +\text{h.c.})
   \label{2}
 \end{equation}
 and
 \begin{equation}
   \mathcal{L}_\text{Q}=i{Q_1^\dag} \bar{\sigma}^\mu {D_\mu }Q_1 + i{Q_2^\dag} \bar{\sigma}^\mu {D_\mu }Q_2 - ({m_Q} b_{ij} Q^i_1 Q^j_2 + \text{h.c.}),
   \label{3}
 \end{equation}
 where $D_\mu$ is the covariant derivative $D_\mu \equiv \partial_\mu-igW^a_\mu \tau^a-ig'B_\mu Y$ with $\tau^as$ being the generators for the corresponding $\text{SU}(2)_\text{L}$ representations.
 The constants $a_{ij}$ and $b_{ij}$ render the gauge invariance of the $a_{ij} T^i T^j$ and $b_{ij} Q^i_1 Q^j_2$ terms, and can be decoded from Clebsch-Gordan (CG) coefficients multiplied by a factor to normalize the mass terms. The nonzero values are 
\begin{align}
   a_{13}&=a_{31}=\frac{1}{2},  ~a_{22}=-\frac 12;\\
   b_{14}&=b_{32}=1, ~b_{23}=b_{41}=-1.
 \end{align}
The Yukawa interactions between the dark sector multiplets and the Higgs doublet are given by
 \begin{equation}
   \mathcal{L}_\text{Yukawa}= y_1 c_{ijk} Q^i_1 T^j H^k +  y_2 d_{ijk} Q^i_2 T^j \tilde{H}^k + \text{h.c.},
   \label{4}
 \end{equation}
where $H$ is the Higgs doublet and $\tilde{H}\equiv i\sigma^2 H^*$ expressed by
 \begin{equation}
   H=\left( \begin{array}{c}
     {H^+}\\
     {H^0}
   \end{array} \right)
   \in (\bm{2},\frac 12)
   , \hspace{0.5cm}
   \tilde{H}=i \begin{pmatrix}
   0 & -i \\
   i & 0
 \end{pmatrix} \left( \begin{array}{c}
   {H^{-}}\\
   {H^{0*}}
 \end{array} \right)
 =\left( \begin{array}{c}
   {H^{0*}}\\
   {-H^-}
 \end{array} \right)
 \in (\bm{2},-\frac 12).
   \label{6}
 \end{equation}
 After the EWSB, the Higgs doublet obtains a vacuum expectation value $v$, and can be written in the unitary gauge as
 \begin{equation}
   H=\frac {1}{\sqrt{2}} \left( \begin{array}{c}
     {0}\\
     {v+h}
   \end{array} \right).
   \label{7}
 \end{equation}
 In the same way, by using the CG coefficients we can deduce the nonzero valus of $c_{ijk}$ and $d_{ijk}$ are
\begin{align}
  c_{132}&=-\frac{1}{\sqrt{2}}, ~c_{222}=-\frac{1}{\sqrt{3}}, ~c_{312}=\frac {1}{\sqrt{6}};\\
  d_{231}&=\frac {1}{\sqrt{6}}, ~d_{321}=\frac{1}{\sqrt{3}}, ~d_{411}=-\frac{1}{\sqrt{2}}.
 \end{align}
Thus the explicit Yukawa interactions can be written as
\begin{equation} \begin{split}
  \mathcal{L}_\text{Yukawa} &=y_1 (v+h) \left(\frac {1}{\sqrt{6}}Q^-_1T^+ - \frac {1}{\sqrt{3}}Q^0_1T^0 - \frac {1}{\sqrt{2}} Q^+_1T^-\right)\\
  \quad  &+ y_2 (v+h) \left(-\frac {1}{\sqrt{2}}Q^-_2T^+ + \frac {1}{\sqrt{3}}Q^0_2T^0 + \frac{1}{\sqrt{6}}Q^+_2T^-\right) + \text{h.c.}
  \label{8}
\end{split} \end{equation}
Then we rewrite all the mass terms into a matrix form:
\begin{align}
   \mathcal{L}_\text{mass} &= -\frac 12
   \left( \begin{array}{ccc}
       {T^0} & {Q_1^0} & {Q_2^0}
   \end{array} \right)
   \mathcal{M}_\text{N}
    \left( \begin{array}{c}
       {T^0}\\
       {Q_1^0}\\
       {Q_2^0}
     \end{array} \right)
   - \left( \begin{array}{ccc}
       {T^-} & {Q_1^-} & {Q_2^-}
   \end{array} \right)
   \mathcal{M}_\text{C}
    \left( \begin{array}{c}
       {T^+}\\
       {Q_1^+}\\
       {Q_2^+}
     \end{array} \right)
     -m_\text{Q} Q^{--}_1Q^{++}_2 + \text{h.c.}\notag\\
     &= -\frac 12 \sum_{i=1}^3 m_{\chi _i^0} {\chi _i^0} {\chi _i^0} - \sum_{i=1}^3 m_{\chi _i^\pm} {\chi _i^-} {\chi _i^+} - m_{\chi^{\pm\pm}} {\chi^{--}} {\chi^{++}} + \text{h.c.},
 \end{align}
 where $\chi _i^0$, $\chi _i^\pm$, and $\chi^{\pm\pm}$ are mass eigenstates, and $m_{\chi _i^0}$ , $m_{\chi _i^\pm}$, and $m_{\chi^{\pm \pm}}$ are the masses of corresponding mass eigenstates.
 The mass matrixes $\mathcal{M}_\text{N}$ and $\mathcal{M}_\text{C}$ are given by
\begin{equation}
  \mathcal{M}_\text{N}= \begin{pmatrix}
    m_T & \frac{1}{\sqrt 3} y_1 v & -\frac{1}{\sqrt 3} y_2 v \\
    \frac{1}{\sqrt 3} y_1 v & 0 & m_Q \\
    -\frac{1}{\sqrt 3} y_2 v & m_Q & 0
  \end{pmatrix}
  ,
  \mathcal{M}_\text{C}= \begin{pmatrix}
    m_T & \frac{1}{\sqrt 2} y_1 v & -\frac{1}{\sqrt 6} y_2 v \\
    -\frac{1}{\sqrt 6} y_1 v & 0 & -m_Q \\
    \frac{1}{\sqrt 2} y_2 v & -m_Q & 0
  \end{pmatrix}
  ,
  m_{\chi^{\pm \pm}}=m_Q.
  \label{6}
\end{equation}
We can use three unitary matrixes $\mathcal{N}$, $\mathcal{C}_\text{L}$ and $\mathcal{C}_\text{R}$ to diagonalize $\mathcal{M}_\text{N}$ and $\mathcal{M}_\text{C}$:
\begin{equation}
  \mathcal{N}^T\mathcal{M}_\text{N}\mathcal{N}=\text{diag}(m_{\chi _1^0},m_{\chi _2^0},m_{\chi _3^0}),\hspace{0.5cm}
  \mathcal{C}_\text{R}^T\mathcal{M}_\text{C}\mathcal{C}_\text{L}=\text{diag}(m_{\chi_1^\pm},m_{\chi_2^\pm},m_{\chi_3^\pm}).
  \label{10}
\end{equation}
Thus, the mass eigenstates are linked to the gauge eigenstates via
\begin{equation}
    \left( \begin{array}{c}
       {T^0}\\
       {Q_1^0}\\
       {Q_2^0}
     \end{array} \right)
     =\mathcal{N} \left( \begin{array}{c}
       {\chi _1^0}\\
       {\chi_2^0}\\
       {\chi _3^0}
     \end{array} \right),\hspace{0.5cm}
    \left( \begin{array}{c}
       {T^+}\\
       {Q_1^+}\\
       {Q_2^+}
     \end{array} \right)
     =\mathcal{C}_L \left( \begin{array}{c}
       {\chi _1^+}\\
       {\chi_2^+}\\
       {\chi _3^+}
     \end{array} \right),\hspace{0.5cm}
    \left( \begin{array}{c}
       {T^-}\\
       {Q_1^-}\\
       {Q_2^-}
     \end{array} \right)
     =\mathcal{C}_R \left( \begin{array}{c}
       {\chi _1^-}\\
       {\chi_2^-}\\
       {\chi _3^-}
     \end{array} \right).
  \label{11}
\end{equation}
For convenience, we adopt the mass orders $0 \le m_{\chi _1^0} \le m_{\chi _2^0} \le m_{\chi _3^0}$ and $0 \le m_{\chi _1^\pm} \le m_{\chi _2^\pm} \le m_{\chi _3^\pm}$,
which can be realized by adjusting the $\mathcal{N}$, $\mathcal{C}_\text{L}$ and $\mathcal{C}_\text{R}$. 
Because of the discrete $Z_2$ symmetry, the lightest Majorana fermion $\chi _1^0$ would be the DM candidate if it is the lightest dark sector fermion. 
Note that we will not consider any CP-violation phase in this work, and thus there are only four independent parameters in the TQDM model: $m_T$, $m_Q$, $y_1$ and $y_2$.

Besides, we can construct 4-component Dirac spinors from 2-component Weyl spinors:
\begin{equation}
  \Psi_i=\left( \begin{array}{c}
    {\chi _i^0}\\
    ({\chi_i^{0}})^\dagger
  \end{array} \right),\hspace{0.5cm}
  X^-_i=\left( \begin{array}{c}
    {\chi _i^-}\\
    ({\chi_i^+})^\dagger
  \end{array} \right),\hspace{0.5cm}
  X^{--}=\left( \begin{array}{c}
    {\chi^{--}}\\
    ({\chi^{++}})^\dagger
  \end{array} \right).
  \label{12}
\end{equation}
The mass and kinetic terms can be rewritten as
\begin{align}
  \mathcal{L}_\text{mass} &=-\frac 12 m_{\chi^0_i} \overline{\Psi_i}\Psi_i - m_{\chi_i^\pm} \overline{X_i^-}X_i^- -m_{\chi^{\pm\pm}} \overline{X^{--}}X^{--},\\
  \mathcal{L}_\text{kinetic} &=\frac 12 i \overline{\Psi_i}\gamma^\mu\partial_\mu\Psi_i + i \overline{X_i^-}\gamma^\mu\partial_\mu X_i^- + i \overline{X^{--}}\gamma^\mu\partial_\mu X^{--}.
\end{align}
The Yukawa interaction terms are
\begin{align}
  \mathcal{L}_\text{Yukawa}&=
  \left(-\frac{y_1}{\sqrt{2}}\mathcal{C}_{\text{L},2i}\mathcal{C}_{\text{R},1j}+\frac{y_1}{\sqrt{6}}\mathcal{C}_{\text{L},1i}\mathcal{C}_{\text{R},2j}+\frac{y_2}{\sqrt{6}}\mathcal{C}_{\text{L},3i}\mathcal{C}_{\text{R},1j}-\frac{y_2}{\sqrt{2}}\mathcal{C}_{\text{L},1i}\mathcal{C}_{\text{R},3j}\right)\overline{X_i^-} P_\text{L} X_j^-h\notag\\
  &+ \left(-\frac{y_1}{\sqrt{2}}\mathcal{C}^*_{\text{R},1i}\mathcal{C}^*_{\text{L},2j}+\frac{y_1}{\sqrt{6}}\mathcal{C}^*_{\text{R},2i}\mathcal{C}^*_{\text{L},1j}+\frac{y_2}{\sqrt{6}}\mathcal{C}^*_{\text{R},1i}\mathcal{C}^*_{\text{L},3j}-\frac{y_2}{\sqrt{2}}\mathcal{C}^*_{\text{R},3i}\mathcal{C}^*_{\text{L},1j}\right)\overline{X_i^-} P_\text{R} X_j^-h \label{higgs}\\
  &-\left(\frac {y_1}{\sqrt{3}}\mathcal{N}_{2i}\mathcal{N}_{1j}-\frac {y_2}{\sqrt{3}}\mathcal{N}_{3i}\mathcal{N}_{1j}\right)\overline{\Psi_i}P_\text{L}\Psi_jh
  -\left(\frac {y_1}{\sqrt{3}}\mathcal{N}^*_{1i}\mathcal{N}^*_{2j}-\frac {y_2}{\sqrt{3}}\mathcal{N}^*_{1i}\mathcal{N}^*_{3j}\right)\overline{\Psi_i}P_\text{R}\Psi_jh.\notag
\end{align}
The interaction terms with the photon are
\begin{equation}
  \mathcal{L}_\text{photon} =-e A_\mu \overline{X_i^-}\gamma^\mu X^-_i - 2e A_\mu \overline{X^{--}}\gamma^\mu X^{--}.
  \label{17}
\end{equation}
The interaction terms with the $Z$ boson are
\begin{align}
  \mathcal{L}_\text{Z}&=\left(-c_\text{W}g\mathcal{C}^*_{\text{R},1i}\mathcal{C}_{\text{R},1j}+\frac {g(s_\text{W}^2-c_\text{W}^2)}{2c_\text{W}} \mathcal{C}^*_{\text{R},2i}\mathcal{C}_{\text{R},2j} -\frac {g(3c_\text{W}^2+s_\text{W}^2)}{2c_\text{W}}\mathcal{C}^*_{\text{R},3i}\mathcal{C}_{\text{R},3j}\right)\overline{X_i^-}\gamma^\mu P_\text{L} X_j^-Z_\mu\notag\\
  &+\left(-c_\text{W}g\mathcal{C}_{\text{L},1i}\mathcal{C}^*_{\text{L},1j}-\frac {g(3c_\text{W}^2+s_\text{W}^2)}{2c_\text{W}} \mathcal{C}_{\text{L},2i}\mathcal{C}^*_{\text{L},2j} +\frac {g(s_\text{W}^2-c_\text{W}^2)}{2c_\text{W}}\mathcal{C}_{\text{L},3i}\mathcal{C}^*_{\text{L},3j}\right)\overline{X_i^-}\gamma^\mu P_\text{R} X_j^-Z_\mu\notag\\
  &+\frac {g}{4c_\text{W}}(\mathcal{N}^*_{2i}\mathcal{N}_{2j}-\mathcal{N}^*_{3i}\mathcal{N}_{3j})\overline{\Psi_i}\gamma^\mu P_\text{L}\Psi_j Z_\mu -\frac {g}{4c_\text{W}}(\mathcal{N}_{2i}\mathcal{N}^*_{2j}-\mathcal{N}_{3i}\mathcal{N}^*_{3j})\overline{\Psi_i}\gamma^\mu P_\text{R}\Psi_j Z_\mu\\
  &+\frac {g(s_\text{W}^2-3c_\text{W}^2)}{2c_\text{W}}\overline{X^{--}}\gamma^\mu X^{--}Z_\mu.\notag
\end{align}
Here $s_\text{W}\equiv\sin\theta_\text{W}$ and $c_\text{W}\equiv\cos\theta_\text{W}$ with $\theta_\text{W}$ denoting the Weinberg angle. Finally, the interaction terms with the $W$ boson are
\begin{align}
  \mathcal{L}_\text{W} &=
  g\left(-\mathcal{N}^*_{1i}\mathcal{C}_{\text{R},1j}+\sqrt{2}\mathcal{N}^*_{2i}\mathcal{C}_{\text{R},2j}+\frac{\sqrt{6}}{2}\mathcal{N}^*_{3i}\mathcal{C}_{\text{R},3j}\right)\overline{\Psi_i}\gamma^\mu P_\text{L} X^-_j W^+_\mu\notag\\
  &-g\left(\mathcal{N}_{1i}\mathcal{C}^*_{\text{L},1j}+\frac{\sqrt{6}}{2}\mathcal{N}_{2i}\mathcal{C}^*_{\text{L},2j}+{\sqrt{2}}\mathcal{N}_{3i}\mathcal{C}^*_{\text{L},3j}\right)\overline{\Psi_i}\gamma^\mu P_\text{R} X^-_j W^+_\mu\notag\\
  &+g\left(-\mathcal{C}^*_{R,1i}\mathcal{N}_{1j}+\sqrt{2}\mathcal{C}^*_{\text{R},2i}\mathcal{N}_{2j}+\frac{\sqrt{6}}{2}\mathcal{C}^*_{\text{R},3i}\mathcal{N}_{3j}\right)\overline{X^-_i}\gamma^\mu P_\text{L} \Psi_j W^-_\mu\notag\\ \label{w}
  &-g\left(\mathcal{C}_{\text{L},1i}\mathcal{N}^*_{1j}+\frac{\sqrt{6}}{2}\mathcal{C}_{\text{L},2i}\mathcal{N}^*_{2j}+{\sqrt{2}}\mathcal{C}_{\text{L},3i}\mathcal{N}^*_{3j}\right)\overline{X^-_i}\gamma^\mu P_\text{R} \Psi_j W^-_\mu\\
  &+\frac{\sqrt{6}}{2}g\mathcal{C}^*_{\text{R},2i}\overline{X^-_i}\gamma^\mu P_\text{L} X^{--}W^+_\mu-\frac{\sqrt{6}}{2}g\mathcal{C}_{\text{L},3i}\overline{X^-_i}\gamma^\mu P_\text{R} X^{--}W^+_\mu\notag\\
  &+\frac{\sqrt{6}}{2}g\mathcal{C}_{\text{R},2i}\overline{X^{--}}\gamma^\mu P_\text{L} X^{-}_iW^-_\mu-\frac{\sqrt{6}}{2}g\mathcal{C}^*_{\text{L},3i}\overline{X^{--}}\gamma^\mu P_\text{R} X^{-}_iW^-_\mu.\notag
\end{align}

\subsection{Mass spectrum}
\label{mass_spectrum}

Masses of dark sector fermions and are determined by the parameter set ($m_T$, $m_Q$, $y_1$, $y_2$).
As the mass spectrum significantly affects the kinematics of their production and decay processes at colliders,
we carry out a careful calculation for the masses with one-loop corrections.
Details of the calculation are not described in this paper, but interested readers may refer to Refs.~\cite{Tait:2016qbg,Baro:2009gn,Denner:1991kt,Fritzsche:2002bi}.
In some parameter regions with $y_1 y_2 <0$, the condition of $m_{\chi^0_1}<m_{\chi^\pm_1}$ satisfied at tree level may not hold at one-loop level~\cite{Tait:2016qbg}. Such parameter regions should be excluded, since there is no available DM candidate.

There are some symmetries regarding the Yukawa couplings.
If one exchanges the values of $y_1$ and
$y_2$ ($y_1\leftrightarrow y_2$), or simultaneously change the signs of $y_1$ and $y_2$ ($y_1\to -y_1$, $y_2\to -y_2$), the mass spectrum would not change.
Another interesting feature is that the dark sector respects a global custodial symmetry when $y_1=\pm y_2$~\cite{Tait:2016qbg,Cai:2016sjz}.
This custodial symmetry ensures that $m_{\chi^0_i}=m_{\chi^\pm_i}$ ($i=1,2,3$) at tree level.
At one-loop level, the limit $y_1= - y_2$ leads to $m_{\chi^0_1}>m_{\chi^\pm_1}$ and hence is not interested.
On the other hand, one-loop corrections in the limit $y_1= y_2$ lift the masses of charged fermions, resulting in $m_{\chi^\pm_i} - m_{\chi^0_i} \simeq \mathcal{O}(100)~\si{MeV}$.
Such a small mass splitting could give rise to a disappearing track signature at colliders.

\begin{figure}[!t]
\centering
\subfigure[~$m_{\chi^0_1}$ contours\label{mass:a}]
{\includegraphics[width=.42\textwidth,trim={0 15 0 10},clip]{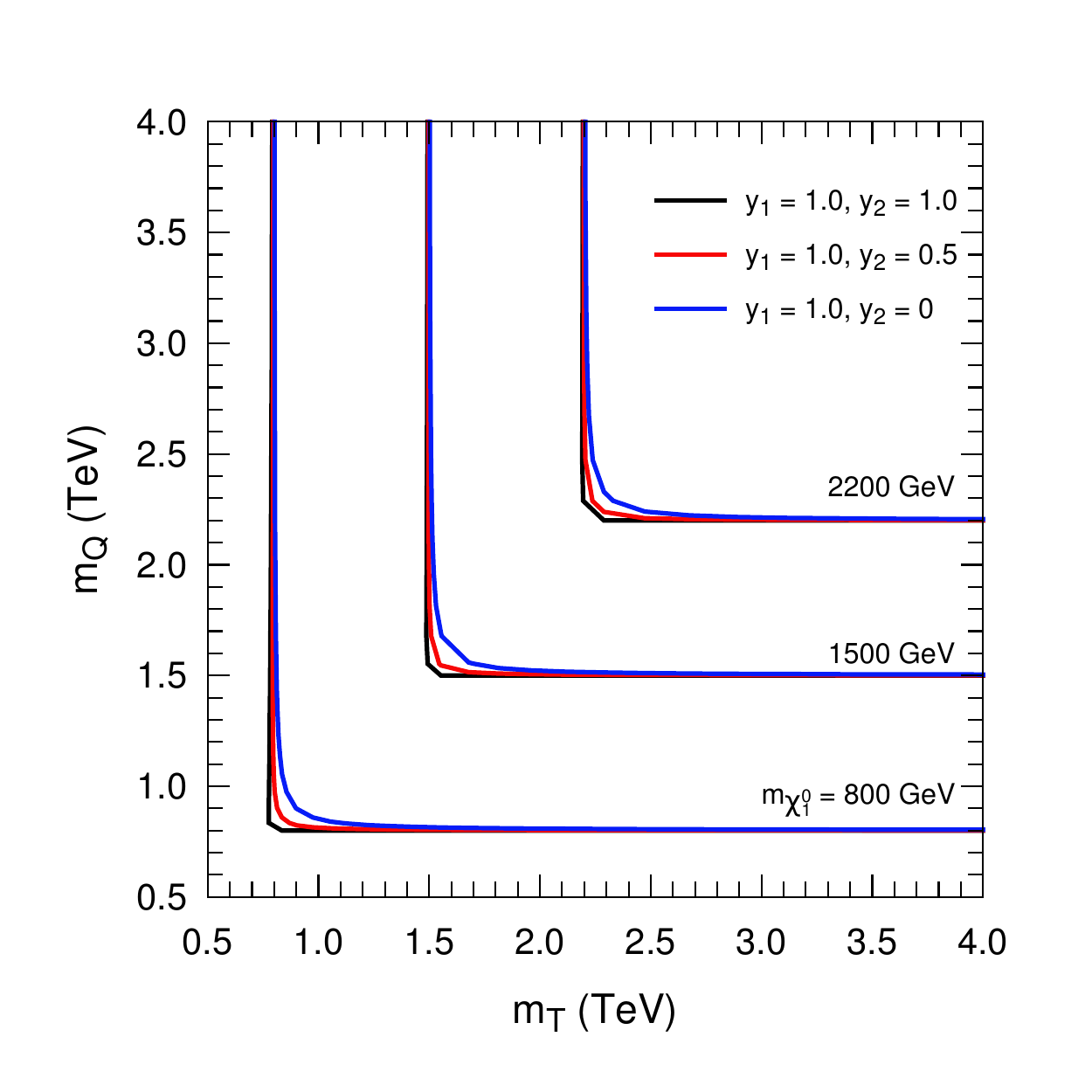}}
\subfigure[~$y_1=1.0$, $y_2=0.7$\label{mass:b}]
{\includegraphics[width=.42\textwidth,trim={0 15 0 10},clip]{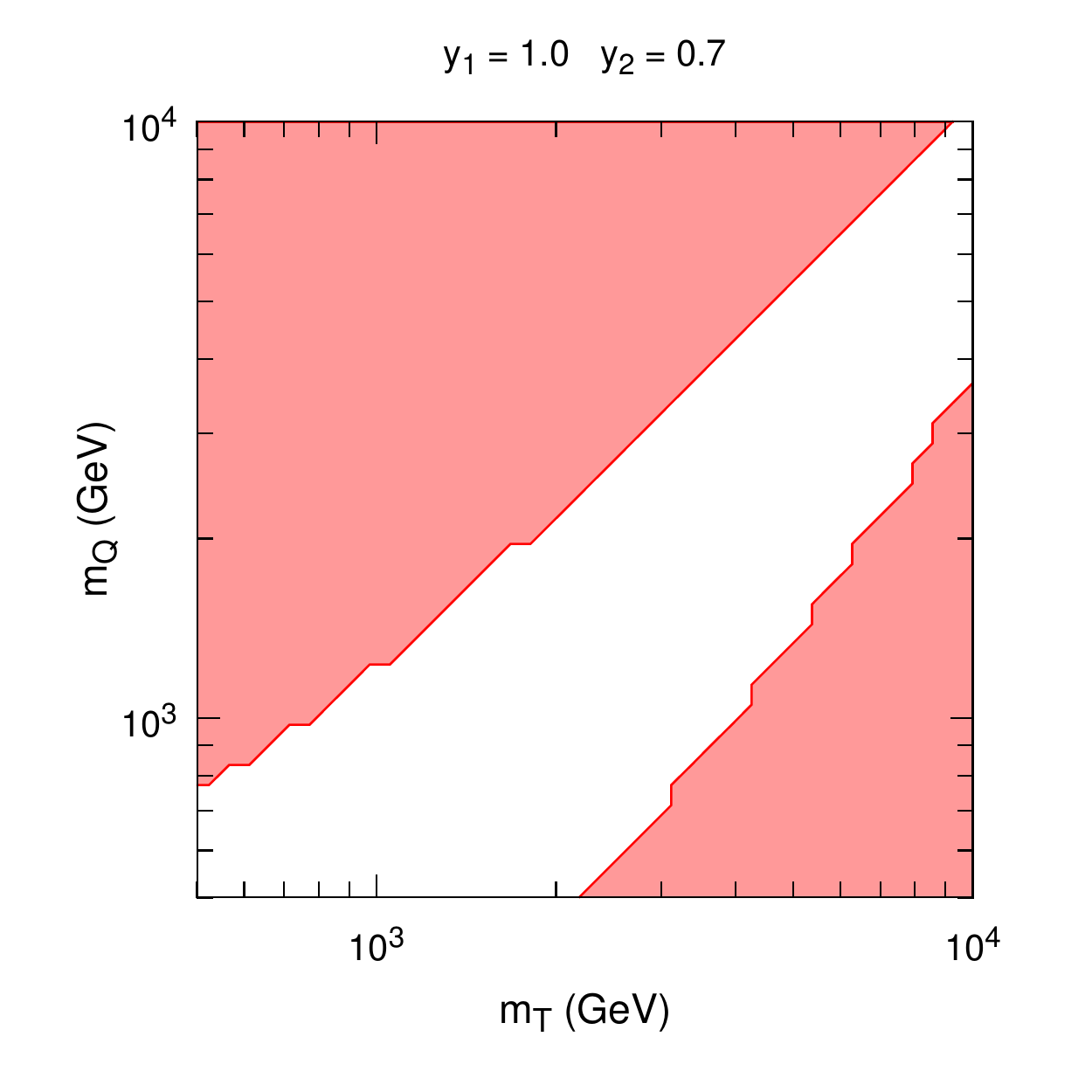}}
\subfigure[~$y_1=0.3$, $y_2=0.7$\label{mass:c}]
{\includegraphics[width=.42\textwidth,trim={0 15 0 10},clip]{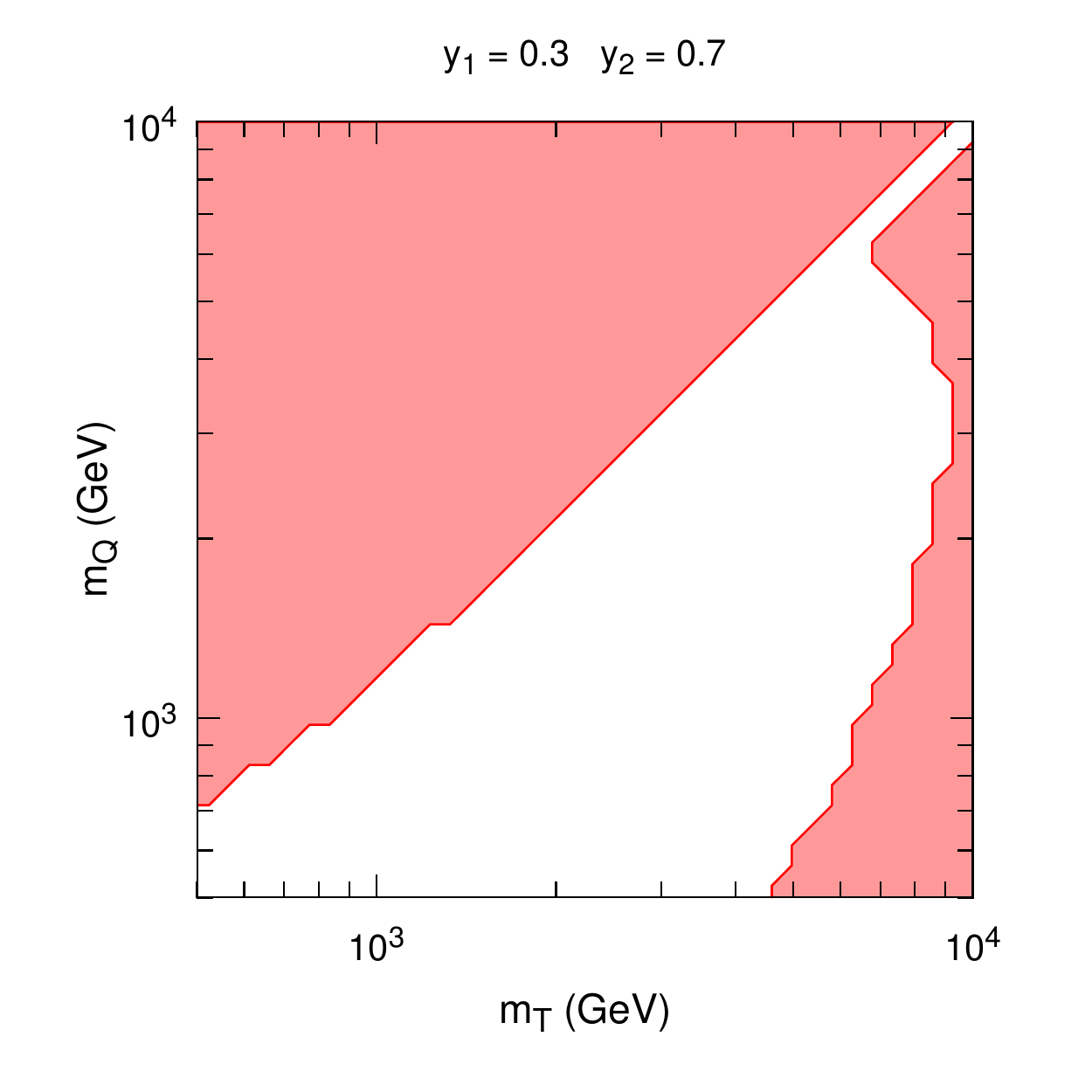}}
\subfigure[~$y_1=1.0$, $y_2=-0.4$\label{mass:d}]
{\includegraphics[width=.42\textwidth,trim={0 15 0 10},clip]{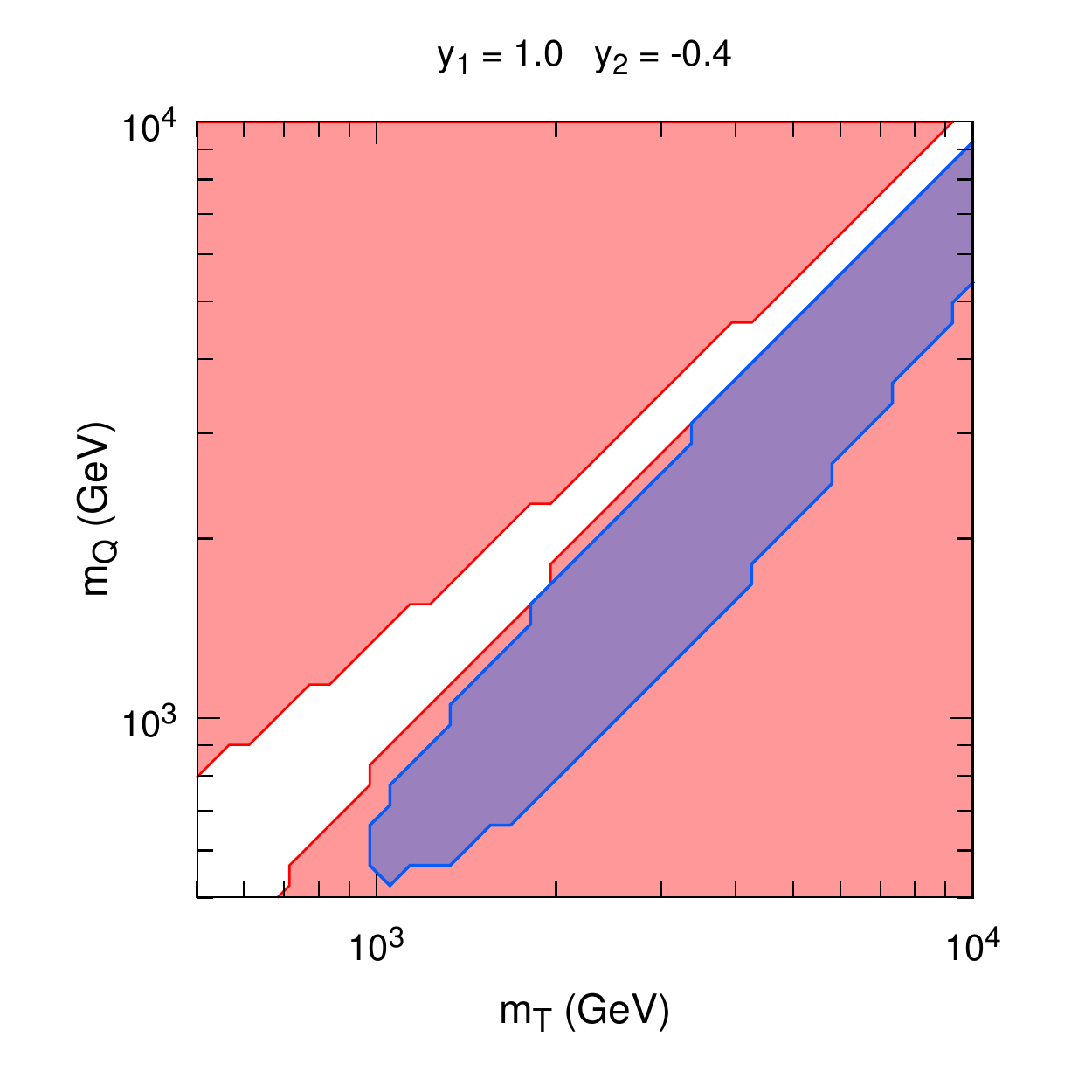}}
\caption{The impacts of $y_1$ and $y_2$ values on $m_{\chi^0_1}$ (a) and the mass splitting $m_{\chi^\pm_1}-m_{\chi^0_1}$ (b), (c), (d) in the $m_T$-$m_Q$ plane.
In the red regions, $m_{\chi^\pm_1}-m_{\chi^0_1}$ is smaller than 250~MeV at one-loop level.
In the blue region, $m_{\chi^\pm_1}-m_{\chi^0_1}$ is negative at one loop-level, and thus $\chi^0_1$ fails to be a DM candidate.}
\label{fig:mass}
\end{figure}

In Fig.~\ref{mass:a}, we show the contours of $m_{\chi^0_1}$ for different sets of $y_1$ and $y_2$ in the $m_T$-$m_Q$ plane. We can see that when $m_T$ and $m_Q$ are $\gtrsim 1~\si{TeV}$, $y_1$ and $y_2$ do not significantly affect the value of $m_{\chi^0_1}$.
This is because the mixing mass terms between the triplet and quadruplets are just determined by the Yukawa interactions at the order of the EWSB scale $\sim \mathcal{O}(100)~\si{GeV}$.
In this case, $m_{\chi^0_1}$ is close to either $m_T$ or $m_Q$, depending on which one is smaller.
Consequently, collider searches for these parameter regions may not be sensitive to $y_1$ and $y_2$, and can set general limits on $m_T$ and $m_Q$.

Figures \ref{mass:b}, \ref{mass:c}, and \ref{mass:d} demonstrate special regions for the mass splitting $m_{\chi^\pm_1}-m_{\chi^0_1}$ with three sets of Yukawa couplings.
The red regions denote the regions where $m_{\chi^\pm_1}-m_{\chi^0_1}~250~\si{MeV}$ are at one-loop level.
In such regions, the disappearing track channel could be quite sensitive.
The blue region in Fig.~\ref{mass:d} corresponds to $m_{\chi^0_1}>m_{\chi^\pm_1}$, which excludes $\chi^0_1$ as a DM candidate.
It can be seen that small mass splittings are quite common in this model, even for the case of $y_1\ne y_2$.
Therefore, it is worthy of considering disappearing track searches at the LHC and future colliders.

\section{LHC and SPPC Searches}
\label{sec:LHC}

In this section, we investigate the current LHC constraints on the TQDM model in several search channels by reinterpreting ATLAS analyses at $\sqrt{s}=13~\si{TeV}$.
We further explore the prospect of the future $pp$ collider SPPC, based on Monte Carlo simulation.
The collision energy and the integrated luminosity of the SPPC are set to be $100~\si{TeV}$ and $3~\si{ab^{-1}}$, respectively.

In our simulation, \texttt{FeynRules~2.3.26}~\cite{Alloul:2013bka} is employed to implement the TQDM model.
Signal and background samples are generated by ~\texttt{MadGraph~5.2.1.2}~\cite{Alwall:2014hca} at parton level.
\texttt{Pythia~6.4.28}~\cite{Sjostrand:2006za} is used to deal with parton showering, hadronization, and decay processes.
Background events are matched up to two additional jets with the MLM matching scheme~\cite{Mangano:2006rw}, while signal events are matched with one jet for simplicity. We have checked that the difference between 1-jet matching and 2-jet matching for signals is negligible.
\texttt{Delphes~3.3.3}~\cite{deFavereau:2013fsa} is utilized to perform a fast detector simulation.

\subsection{$\text{Monojet}+\missET$ channel}

First, we consider the $\text{monojet}+\missET$ channel, where the final state involves an energetic jet and a large missing transverse momentum. This channel is clean and distinctive and has been widely used to search for large extra dimensions~\cite{ArkaniHamed:1998rs}, SUSY models~\cite{Yu:2012kj}, and generic WIMPs~\cite{Beltran:2010ww,Fox:2011pm,Xiang:2015lfa,ATLAS:2017dnw} at the Tevatron and the LHC.
In the TQDM model, pair production of dark sector fermions associated with a hard ISR jet would also give rise to such a $\text{monojet}+\missET$ final state.

SM backgrounds in the $\text{monojet}+\missET$ channel are dominated by $Z(\to\nu\bar{\nu})+\text{jets}$, $W(\to \ell \nu)+\text{jets}$,
$t\bar{t}+\text{jets}$, and $W(\to \ell\nu)W(\to \ell\nu)+\text{jets}$~\cite{Low:2014cba,ATLAS:2017dnw}.
Other contributions, \textit{e.g.}, from top production associated with additional vector bosons can be neglected~\cite{ATLAS:2017dnw}.
We have carefully compared our simulated backgrounds with those given by the ATLAS $\text{monojet}+\missET$ analysis with $\sqrt{s}=13~\si{TeV}$~\cite{Aaboud:2016tnv}, and find that they are almost perfectly matched with each other.
For the signal simulation, we include all the production processes of $pp\to \chi\chi  +\text{jets}$,
where $\chi$ represents any fermion in ($\chi^0_i$, $\chi^\pm_i$, $\chi^{\pm\pm}$).

\begin{table}[!t]
\setlength{\tabcolsep}{.5em}
\renewcommand{\arraystretch}{1.2}
\begin{tabular}{|c|c|c|}
\hline
 & 13~TeV  LHC & 100~TeV SPPC \\
\hline
\multicolumn{3}{|c|}{Reconstruction conditions}\\
\hline
$\pT(j),\  \vert\eta(j)\vert$ & $>30 ~\si{GeV},\  <2.8$ & $>80 ~\si{GeV},\  <2.8$ \\
$\pT(e),\  |\eta(e)|$ & $>20 ~\si{GeV},\  <2.47$ & $>20 ~\si{GeV},\  <2.47$ \\
$\pT(\mu),\  |\eta(\mu)|$ & $>10 ~\si{GeV},\  <2.5$ & $>10 ~\si{GeV},\  <2.5$ \\
\hline
\multicolumn{3}{|c|}{Cut conditions}\\
\hline
$n_\ell$ & $0$ & $0$ \\
$n_j$ & $\le4$ & $\le4$ \\
$\pT(j_1),\  \vert\eta(j_1)\vert$ & $>250 ~\si{GeV},\  <2.4$& $>1500 ~\si{GeV},\  <2.4$ \\
$\Delta \phi (j, \slashed{\vec{p}}_\mathrm{T})$ & $>0.4$ & $>0.4$ \\
$\slashed{E}_\mathrm{T}$ & $>250 - 1000 ~\si{GeV}$ & $>1200 - 2500 ~\si{GeV}$  \\
\hline
\end{tabular}
\caption{Reconstruction and cut conditions in the $\text{monojet}+\missET$ channel used in the ATLAS analysis at the 13~TeV LHC~\cite{ATLAS:2017dnw} and those adopted for estimating the sensitivity at the 100~TeV SPPC.
Here $j$ and $\ell = e,\mu$ represent jet and lepton, respectively.
$j_1$ means the leading jet, \textit{i.e.,} the jet with the largest $\pT$.
$n_\ell$ and $n_j$ are the numbers of reconstructed leptons and jets, respectively.
$\slashed{\vec{p}}_\mathrm{T}$ denotes the missing transverse momentum vector.}
\label{monojet_1}
\end{table}

\begin{figure}[!t]
\centering
\subfigure[~$y_1 = y_2 =0.5$\label{fig:monojet_results_a}]
{\includegraphics[width=0.48\textwidth,trim={0 15 0 10},clip]{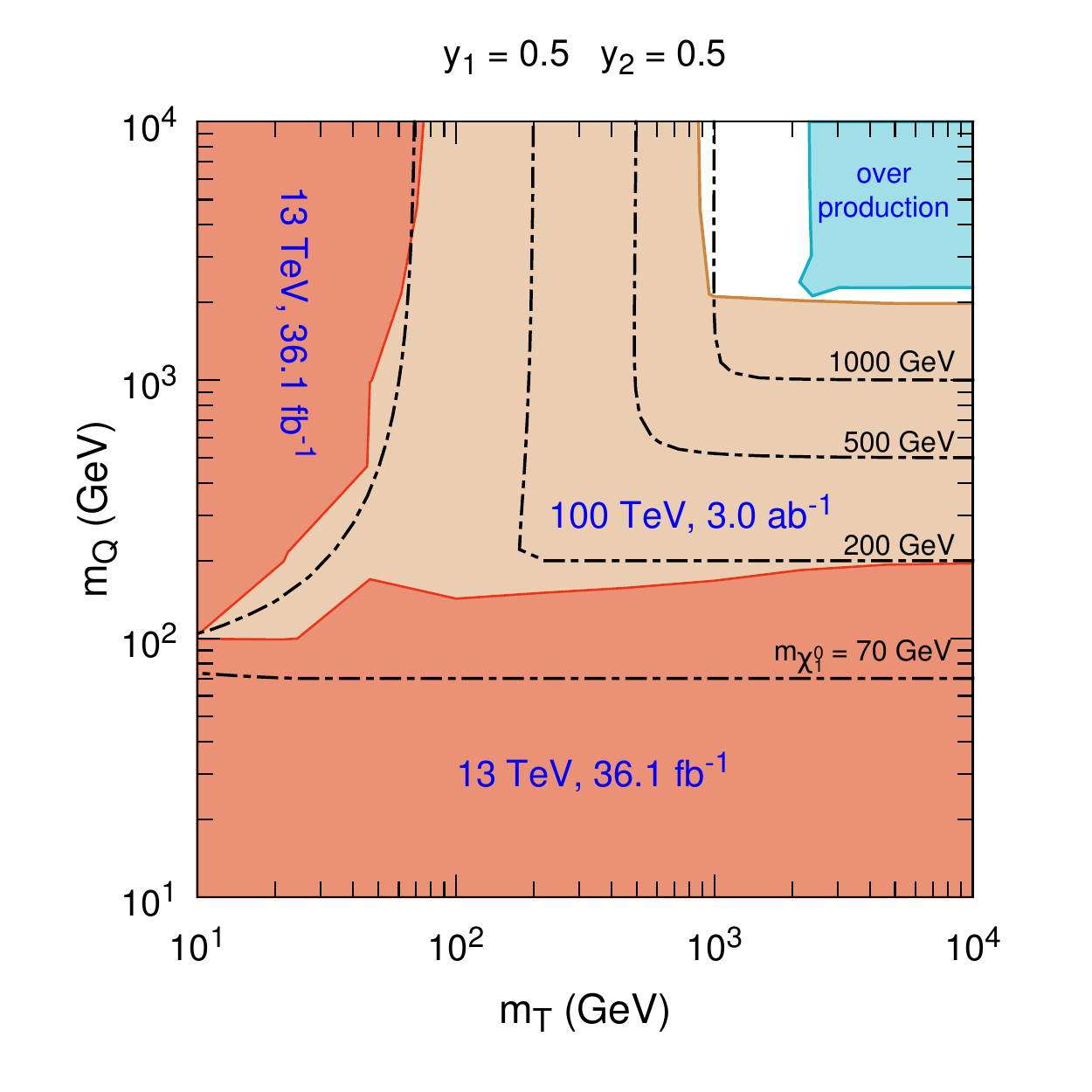}}
\subfigure[~$y_1 = 1.0$, $y_2 =0.5$\label{fig:monojet_results_b}]
{\includegraphics[width=0.48\textwidth,trim={0 15 0 10},clip]{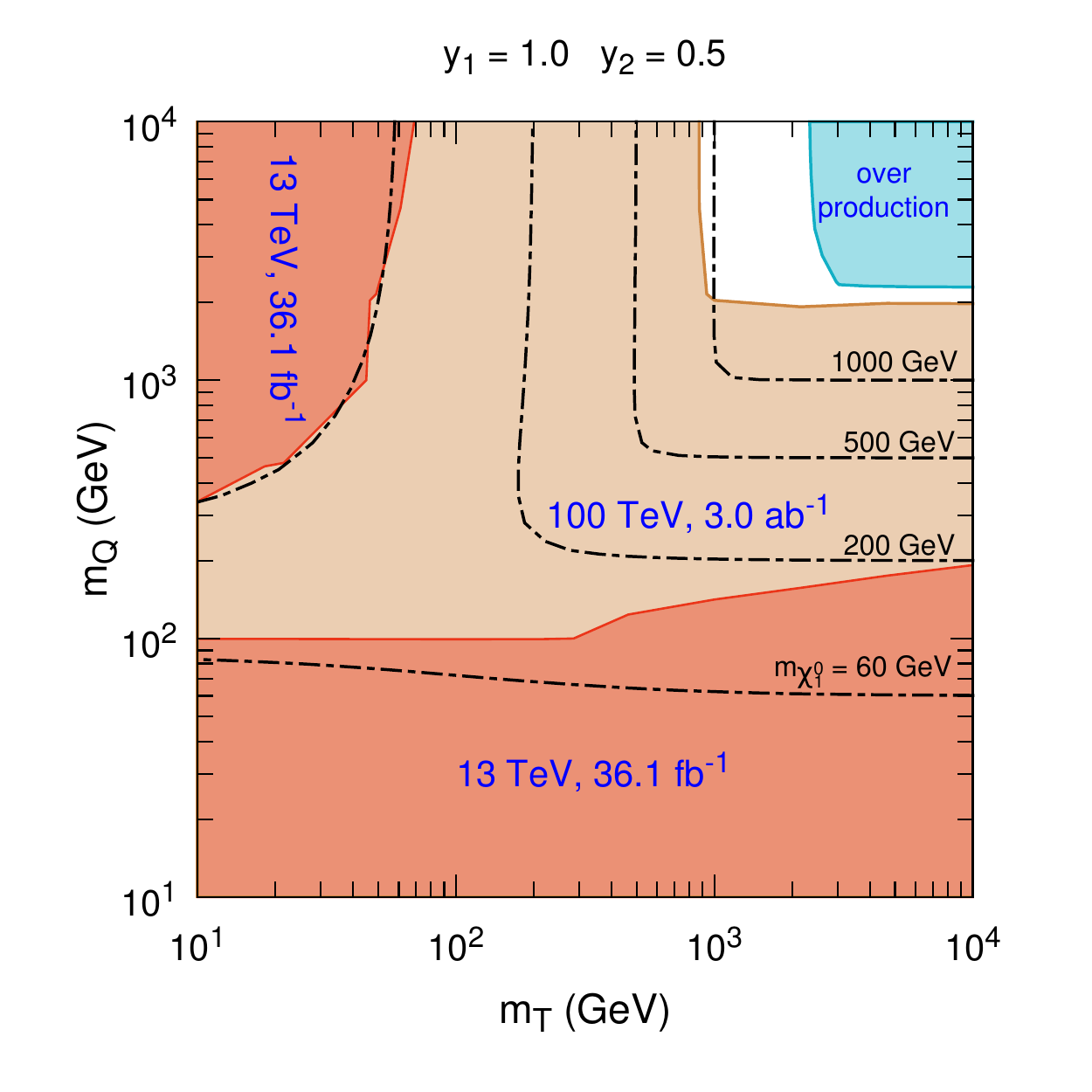}}
\caption{LHC constraints and SPPC sensitivities from the $\text{monojet}+\missET$ channel in the $m_T$-$m_Q$ plane with the fixed Yukawa couplings of $y_1=y_2=0.5$ (a) and $(y_1,y_2)=(1.0,0.5)$ (b).
The orange regions are excluded at $95\%$ C.L. by the ATLAS search at $\sqrt{s}=13~\si{TeV}$ with a data set of $36.1~\si{fb^{-1}}$~\cite{ATLAS:2017dnw}.
The canary yellow regions are expected to be excluded at $95\%$ C.L. at the SPPC with $\sqrt{s}=100~\si{TeV}$ and an integrated luminosity of $3~\si{ab^{-1}}$.
The blue regions represent the parameter regions excluded by overproduction of dark matter in the early universe.
For reading convenience, contours of $m_{\chi^0_1}$ are also demonstrated with the dot-dashed lines.}
\label{fig:monojet_results}
\end{figure}

For evaluating the current LHC constraint, we study the latest result from the ATLAS analysis~\cite{ATLAS:2017dnw} with $\sqrt{s}=13~\si{TeV}$ and an integrated luminosity of 36.1 fb$^{-1}$.
The object reconstruction conditions and kinematic cut conditions used in this analysis are listed in the second column of Table~\ref{monojet_1}.
They require the final state involving at least one energetic central jet, a large $\missET$, and no lepton ($n_\ell=0$ for $\ell=e,\mu$).
We closely follow these cut conditions to reinterpret the experimental result.
The resulting constraint is shown in Fig.~\ref{fig:monojet_results}, where
the orange regions are excluded at 95\% C.L. by the LHC search.

If $m_Q \gg m_T$, $\chi_1^0$ is dominated by the triplet component, and the LHC bound can exclude up to $m_{\chi^0_1} \sim 70~\si{GeV}$.
On the other hand, $m_T \gg m_Q$ leads to a quadruplet-dominated $\chi_1^0$, and the LHC exclusion limit goes up to $m_{\chi^0_1} \sim 200~\si{GeV}$.
This is due to the fact that the pure quadruplets have larger production cross section than the pure triplet, as the quadruplets contain more dark sector particles than the triplet.

Below we estimate the sensitivity of the $\text{monojet}+\missET$ channel at the SPPC with $\sqrt{s}=100~\si{TeV}$.
The signal significance $\mathcal{S}$ can be defined as~\cite{Low:2014cba}
\begin{equation}
  \mathcal{S}=\frac S{\sqrt{B+(\lambda B)^2+(\gamma S)^2}},
  \label{21}
\end{equation}
where $S$ and $B$ are the numbers of signal and background events, respectively, while $\lambda$ and $\gamma$ indicate the systematic uncertainties of the background and the signal, respectively.
In order to improve the significance, one needs to perform some efficient cuts. For the $\text{monojet}+\missET$ channel, cuts on $\missET$ and the transverse momentum of the leading jet ($\pT(j_1)$) are very important.
By analyzing the $\text{monojet}+\missET$ distributions of the backgrounds and some signal benchmark points (BMPs), we can deduce proper values or intervals of the kinematic variables for cut conditions.

\begin{table}[!t]
\setlength{\tabcolsep}{.5em}
\renewcommand{\arraystretch}{1.2}
\begin{tabular}{|c|c|c|c|c|}
\hline
&  $y_1$  &  $y_2$ & $m_T/\si{GeV}$ &  $m_Q/\si{GeV}$ \\
\hline
BMP-a1 & 0.23 & 0.79 & 323  & 473 \\
BMP-a2 & 0.15 & 0.80 & 200  & 700 \\
BMP-a3 & 0.45 & 0.50 & 500  & 600 \\
BMP-a4 & 0.91 & 0.81 & 950  & 860 \\
\hline
\end{tabular}
\caption{Models parameters of the signal BMPs in the $\text{monojet}+\missET$ channel.}
\label{monojet_3}
\end{table}

\begin{figure}[!t]
\centering
\subfigure[~Differential cross sections\label{fig:SPPC_monojet_missEt_a}]
{\includegraphics[width=0.48\textwidth,trim={0 15 0 10},clip]{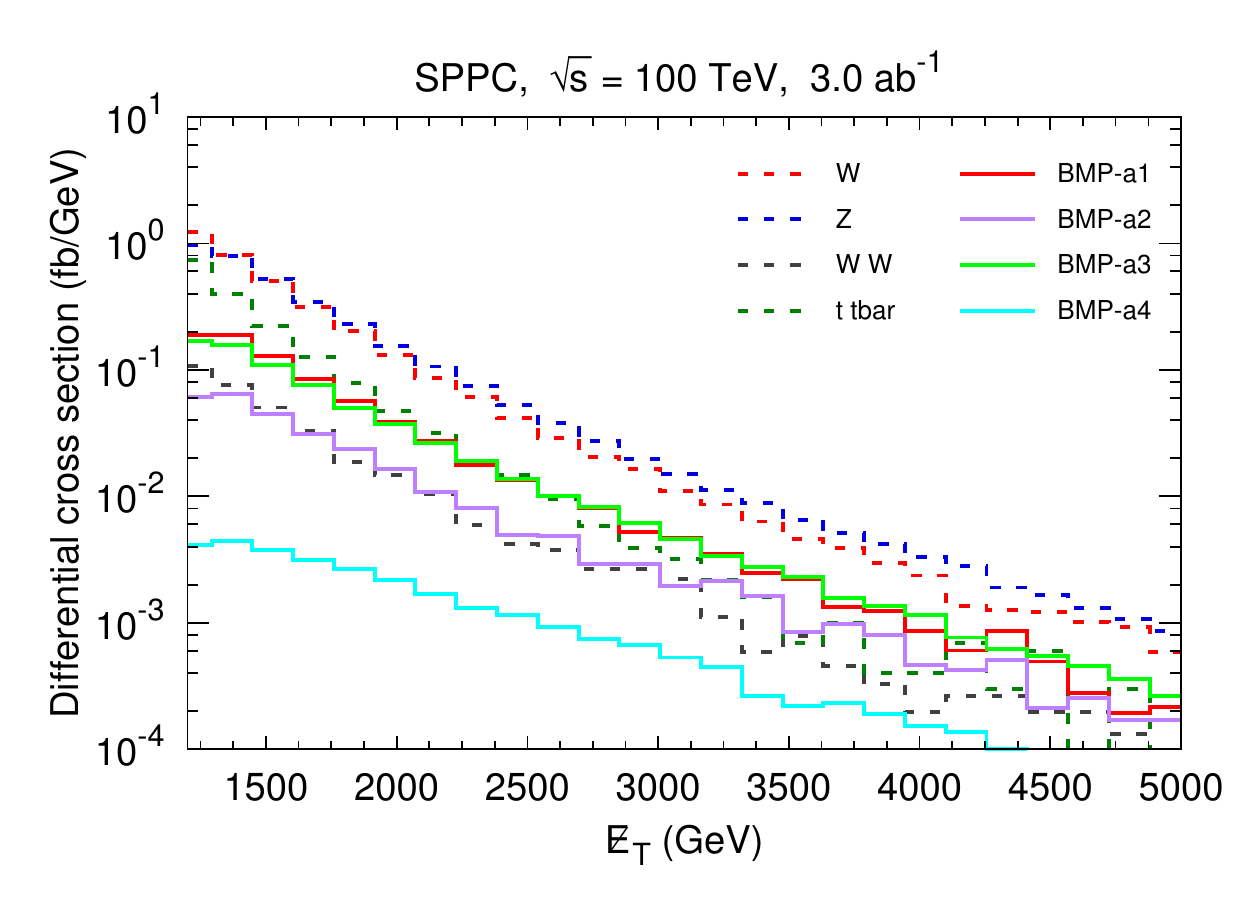}}
\subfigure[~Normalized distributions\label{fig:SPPC_monojet_missEt_b}]
{\includegraphics[width=0.48\textwidth]{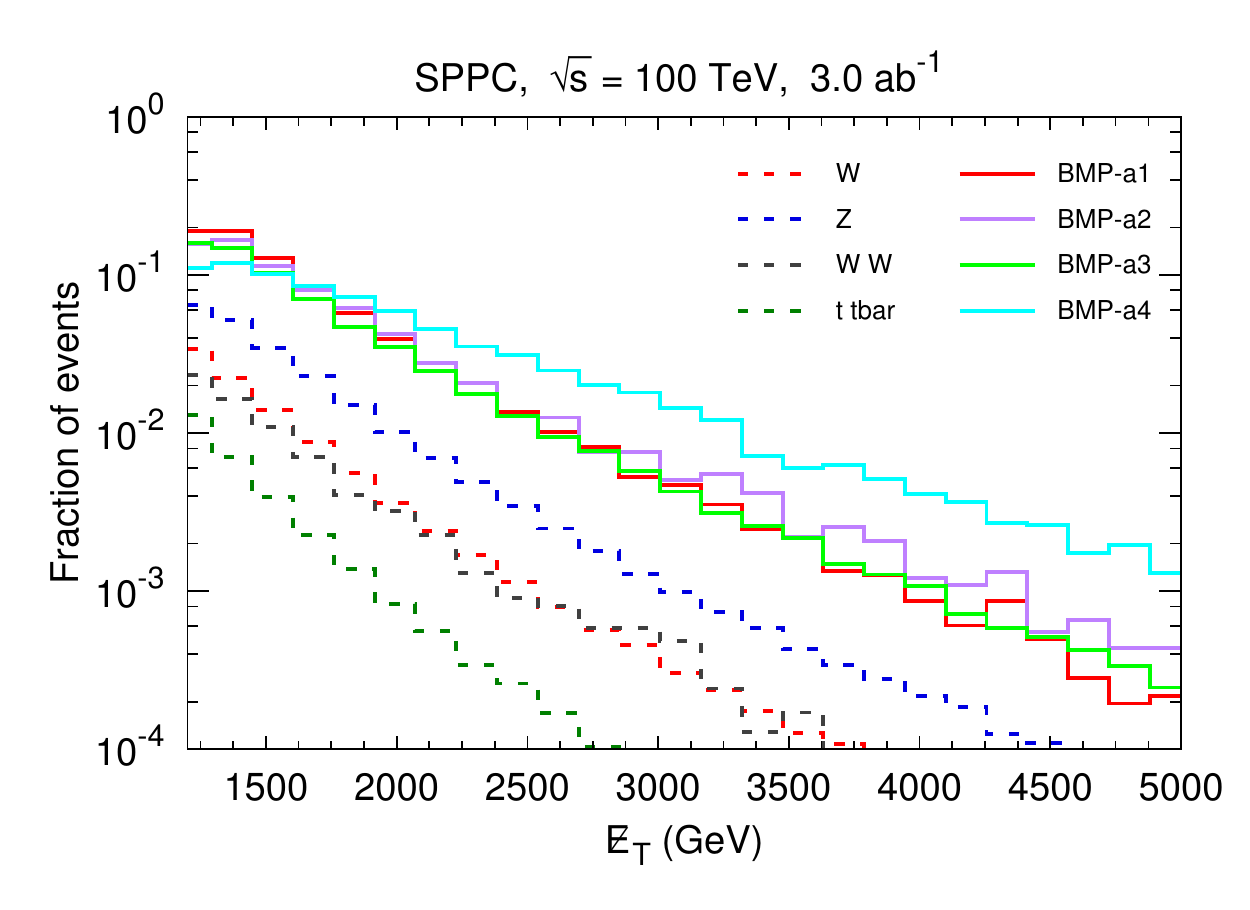}}
\caption{Distributions of differential cross sections (a) and fractions of events (b) as functions of $\slashed{E}_\mathrm{T}$ for backgrounds and signal BMPs in the $\text{monojet}+\missET$ channel at $\sqrt{s}=100~\si{TeV}$.}
\label{fig:SPPC_monojet_missEt}
\end{figure}

We consider four signal BMPs, whose model parameters are tabulated in Table~\ref{monojet_3}.
Figures \ref{fig:SPPC_monojet_missEt_a} and \ref{fig:SPPC_monojet_missEt_b} demonstrate the differential cross sections and the fractions in each bins for backgrounds and BMPs as functions of $\slashed{E}_\mathrm{T}$ at $\sqrt{s}=100~\si{TeV}$.
Here we have performed a preselection by requiring $\pT(j_1)>1200~\si{GeV}$~\cite{Low:2014cba}.
It is obvious that the distributions of the signals are likely to extend to higher $\missET$ than the backgrounds.

The reconstruction and cut conditions we adopt for SPPC are tabulated in the third column of Table~\ref{monojet_1}.
In order to efficiently improve the signal significance, we consider six signal regions with different cuts on $\slashed{E}_\mathrm{T}$, as listed in Table~\ref{monojet_2}.
It is difficult to accurately evaluate the systematic uncertainties for future detectors. Here we simply set $\gamma = 10\%$ and $\lambda = 1\%$ \cite{Low:2014cba}. Although these values may be overly optimistic, they provide a benchmark case for estimating the signal significance.

\begin{table}[!t]
\setlength{\tabcolsep}{.5em}
\renewcommand{\arraystretch}{1.2}
\begin{tabular}{|c|c|c|c|c|c|c|}
\hline
Inclusive signal region & IM1 & IM2 & IM3  & IM4  & IM5  & IM6  \\
\hline
$\slashed{E}_\mathrm{T}$ GeV & $> 1200$ & $> 1400$ & $> 1600$ & $> 1800$  & $> 2000$  & $> 2500$ \\
\hline
\end{tabular}
\caption{Inclusive signal regions with different thresholds of $\slashed{E}_\mathrm{T}$ in the $\text{monojet}+\missET$ channel.}
\label{monojet_2}
\end{table}

95\% C.L. expected exclusion limits of the $\text{monojet}+\missET$ channel at the SPPC with $\sqrt{s}=100~\si{TeV}$ and an integrated luminosity of $3~\si{ab^{-1}}$ are indicated by the canary yellow regions in Fig.~\ref{fig:monojet_results}.
Compared with the capability of the LHC, SPPC will improve the search ranges of the mass parameters by an order of magnitude.
When a pure triplet (quadruplet) $\chi_1^0$, the $\text{monojet}+\missET$ search at the SPPC could reach up to $m_{\chi_1^0}\sim 900~(2000)~\si{GeV}$.

Two Yukawa parameter sets with $y_1=y_2$ and $y_1\neq y_2$, \textit{i.e.}, the cases that the custodial symmetry is respected and violated, are considered in Fig.~\ref{fig:monojet_results_a} and Fig.~\ref{fig:monojet_results_b}. We find that the variation of
the Yukawa couplings does not significantly affect the SPPC sensitivity.
As we have mentioned in Sec.~\ref{mass_spectrum}, this is because Dirac masses induced by $y_1$ and $y_2$ are quite small, compared to TeV-scale $m_T$ and $m_Q$.

In Fig.~\ref{fig:monojet_results}, we also demonstrate the constraints from the observed DM relic density.
We utilize the package \texttt{MadDM~2.0.6}~\cite{Backovic:2013dpa} to calculate the thermal relic density in the TQDM model. All annihilation and coannihilation processes at the freeze-out epoch have been taken into account.
The blue regions in Fig.~\ref{fig:monojet_results} are excluded due to DM overproduction, \textit{i.e.,} the predicted $\Omega_\chi h^2$ is larger than the observed value 0.1186~\cite{Ade:2015xua}.
Compared with this constraint, we can see that the SPPC search will be able to explore a very large region in the allowed parameter space.

\subsection{Disappearing track channel}

In some supersymmetric models, such as the anomaly mediated supersymmetry breaking (AMSB) scenario \cite{Giudice:1998xp,Randall:1998uk},
the lightest chargino $\tilde\chi^\pm_1$ is nearly mass degenerate with the lightest supersymmetric particle (LSP). The lifetime of chargino can be long enough to travel a distinct distance in the inner detector before decaying into the LSP and a very soft, unlikely detected SM particles, such as pion. Therefore, the track arising from such a long-lived chargino seems to disappear, and only leaves hits in the innermost layers.
There would be no hit in the portions of the detector at higher radii, because the LSP passes through the detector without interaction~\cite{ATLAS:2017bna}. This is the reason why such a signature is called disappearing tracks.

In the TQDM model, there are three cases that could lead to a mass degeneracy between $\chi^\pm_1$ and $\chi^0_1$ with $m_{\chi^\pm_1} - m_{\chi^0_1} \sim \mathcal{O}(100)~\si{MeV}$. They would also induce a disappearing tack signal at colliders.
As mentioned in Sec.~\ref{sec:model}, a custodial symmetry in the limit of $|y_1|=|y_2|$ would lead to such a mass spectrum.
Moreover, when $m_T \gg \max(m_Q, |y_1|v, |y_2|v)$ or $m_Q \gg \max(m_T, |y_1|v, |y_2|v)$, the quadruplet components or the triplet component in $\chi^0_1$ is almost dominant.
In these two cases, $\chi^0_1$ is almost degenerate with $\chi^\pm_1$ in mass even for $y_1 \neq y_2$, because the mixing terms in mass matrices are suppressed by $m_T$ or $m_Q$.
In the following study, we focus on the latter two cases with $m_T\ll m_Q$ (pure triplet case) and $m_T \gg m_Q$ (pure quadruplet case) and  take $y_1=y_2=0$ for simplicity.

\begin{table}[!t]
\setlength{\tabcolsep}{.5em}
\renewcommand{\arraystretch}{1.2}
\begin{tabular}{|c|c|c|}
\hline
 & 13 TeV LHC & 100 TeV SPPC\\
\hline
\multicolumn{3}{|c|}{Reconstruction conditions}\\
\hline
$\pT(j),\  \vert\eta(j)\vert$ & $>20 ~\si{GeV},\  <2.8$ & $>80 ~\si{GeV},\  <2.8$ \\
$\pT(e),\  |\eta(e)|$ & $>10 ~\si{GeV},\  <2.47$ & $>10 ~\si{GeV},\  <2.47$ \\
$\pT(\mu),\  |\eta(\mu)|$ & $>10 ~\si{GeV},\  <2.7$ & $>10 ~\si{GeV},\  <2.7$ \\
Tracklet candidate $\pT,\  \vert\eta\vert$ & $>20 ~\si{GeV},\  <2.2$ & $>80 ~\si{GeV},\  <2.2$ \\
\hline
\multicolumn{3}{|c|}{Cut conditions}\\
\hline
$n_{\ell}$ & $0$ & $0$ \\
$n_{j}$ & $\ge1$ & $\ge1$ \\
$n_\text{tracklet}$ & $1$ & $1$ \\
$\pT(j_1)$, $|\eta(j_1)|$ & $>140 ~\si{GeV}$, - & $>1500 ~\si{GeV},\  <2.4$ \\
$\Delta \phi (j_{[\pT>50 ~\si{GeV},\  \text{top4}]}, \slashed{\vec{p}}_\mathrm{T})$ & $>1$ & - \\
$\Delta \phi (j_{[\pT>300 ~\si{GeV},\  \text{top4}]}, \slashed{\vec{p}}_\mathrm{T})$ & - & $>1$ \\
Tracklet $\pT,\  \vert\eta\vert$ & $>100 ~\si{GeV},\ (0.1,1.9)$ & $>500 ~\si{GeV},\ (0.1,1.9)$ \\
$\Delta R (\text{tracklet}, j_{[\pT>50 ~\si{GeV}]})$ & $>0.4$ & - \\
$\Delta R (\text{tracklet}, j_{[\pT>300 ~\si{GeV}]})$ & - & $>0.4$ \\
$\slashed{E}_\mathrm{T}$ & $> 140 ~\si{GeV}$ & $ >1500 ~\si{GeV}$  \\
\hline
\end{tabular}
\caption{Reconstruction and cut conditions in the disappearing track channel.
The second column is used in the ATLAS analysis at the 13~TeV LHC~\cite{ATLAS:2017bna}.
The third column is used for evaluating the sensitivity at the 100~TeV SPPC.
$j$ with a subscript ``top4'' means a jet belongs to the four highest-$\pT$ jets. Note that all the terms ``tracklets'' in the table represent the pixel tracklets.}
\label{disappearing track cuts}
\end{table}

The current ATLAS search for disappearing tracks~\cite{ATLAS:2017bna} is based on $36.1~\si{fb}^{-1}$ of data at $\sqrt{s}=13~\si{TeV}$.
The cut conditions are summarized in the second column of Table~\ref{disappearing track cuts}.
A critical object for this search is called a pixel tracklet, which contains at least four pixel-detector hits and with no hits in the strip semiconductor tracker and transition radiation tracker detectors. 
Besides all these pixel tracklets must not belong to any standard track.
By such a definition, pixel tracklets mimics disappearing tracks we are looking for.

\begin{figure}[!t]
\centering
{\includegraphics[width=.48\textwidth,trim={0 12 0 10},clip]{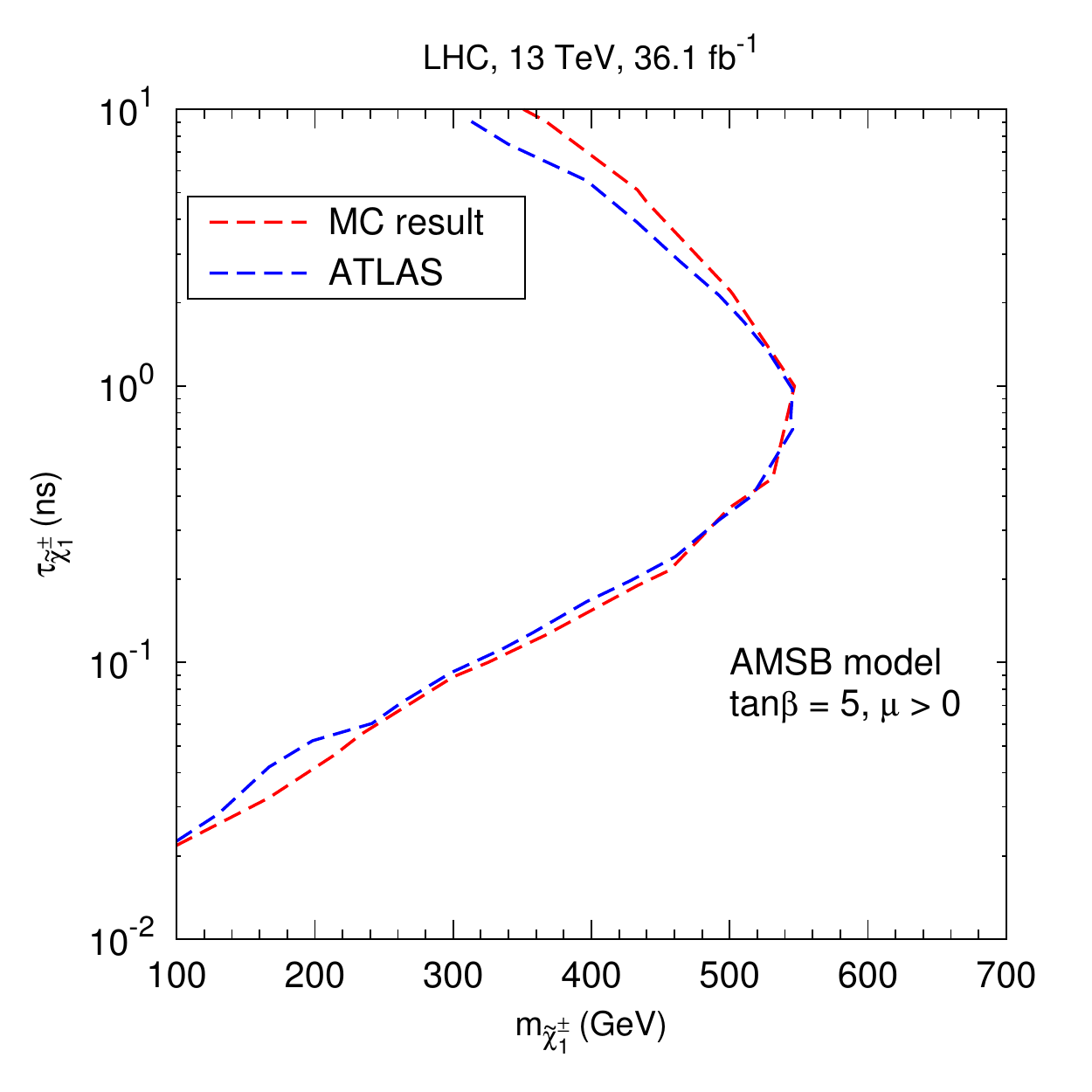}}
\caption{The comparison between the ATLAS result~\cite{ATLAS:2017bna} and our MC result in the disappearing track channel for the $\sigma_\mathrm{vis}$ limit at 95\% C.L. in the $m_{\tilde\chi^\pm_1}$-$\tau_{\tilde\chi^\pm_1}$ plane for the AMSB model.}
\label{disappear track comparison}
\end{figure}

Before analyzing the disappearing track signature in the TQDM model, we should check the validity of our simulation and analysis method.
We attempt to reproduce the $\sigma_\mathrm{vis}$ contour in the $m_{\tilde\chi^\pm_1}$-$\tau_{\tilde\chi^\pm_1}$ plane for the AMSB model according to the $95\%$ C.L. observed limit of $\sigma_\mathrm{vis} = 0.22~\si{fb}$ given in the ATLAS analysis. Here $\sigma_\mathrm{vis}$ is the signal visible cross section after imposing kinematic cuts.
From Fig.~\ref{disappear track comparison}, we can see that our MC result matches with the ATLAS result quit well.
It is worth noting that the disappearing condition of
this latest ATLAS analysis is different from previous works~\cite{Aad:2013yna,Low:2014cba}.
The previous ATLAS analysis~\cite{Aad:2013yna} requires the track length of unstable charged particles larger than $\sim 30~\si{cm}$, while
much shorter tracks, especially the pixel tracklets, are used in the latest analysis.

Then we study the constraint from the disappearing track search on the TQDM model. The signal events are generated through $pp\to \chi\chi +\text{jets}$. A leading jet from ISR is required to ensure a significant $\slashed{E}_\mathrm{T}$, making the trigger more efficient.
For the pure triplet case, the mass splitting between $\chi^\pm_1$ and $\chi^0_1$ is $\sim 167~\si{MeV}$. Therefore, the dominant decay process of $\chi^\pm_1$ is $\chi^\pm_1\to\chi^0_1 \pi^\pm$.
For the pure quadruplets case, $\chi^\pm_1$, $\chi^\pm_2$, and $\chi^{\pm\pm}$ should be all considered.
Because the mass splittings between $\chi^{\pm\pm}$ and $\chi^\pm_1$/$\chi^\pm_2$ are larger than $800~\si{MeV}$,
$\chi^{\pm\pm}$ decays with a lifetime as short as $\sim 10^{-4}~\si{ns}$, and with almost equal branching ratios $\sim 50\%$ to $\chi^\pm_1$ and $\chi^\pm_2$.
Thus, we only need to consider the lifetimes of $\chi^\pm_1$ and $\chi^\pm_2$ for the disappearing track search.

The decay widths of the $\chi^\pm_i \to \chi^0_j \pi^\pm$ processes are given by~\cite{Chen:1999yf}
\begin{eqnarray}
  \Gamma(\chi^\pm_i \to \chi^0_j \pi^\pm) = \frac{f^2_\pi G^2_\mathrm{F}}{4\pi}\frac{|\vec{k}_\pi|}{m^2_{\chi^\pm_i}}\biggl \{\Bigl(O^\mathrm{L}_{ij}+O^\mathrm{R}_{ij}\Bigr)^2\biggl[(m^2_{\chi^\pm_i}-m^2_{\chi^0_j})^2-m^2_\pi(m_{\chi^\pm_i}-m_{\chi^0_j})^2\biggr]\nonumber\\
  + \Bigl(O^\mathrm{L}_{ij}-O^\mathrm{R}_{ij}\Bigr)^2\biggl[(m^2_{\chi^\pm_i}-m^2_{\chi^0_j})^2-m^2_\pi(m_{\chi^\pm_i}+m_{\chi^0_j})^2\biggr]  \biggr \},
  \label{22}
\end{eqnarray}
where $f_\pi\simeq 93~\si{MeV}$ is the pion decay constant, and $G_\mathrm{F}$ is the Fermi coupling constant. 
$O^\mathrm{L}_{ij}$ and $O^\mathrm{R}_{ij}$ are $\chi^0_i \chi^\pm_j W$ couplings and can be read from Eq.~\eqref{w}:
\begin{equation}
  O^\mathrm{L}_{ij}=-\mathcal{N}^*_{1i}\mathcal{C}_{R,1j}+\sqrt{2}\mathcal{N}^*_{2i}\mathcal{C}_{R,2j}+\frac{\sqrt{6}}{2}\mathcal{N}^*_{3i}\mathcal{C}_{R,3j},~~
  O^\mathrm{R}_{ij}=\mathcal{N}_{1i}\mathcal{C}^*_{L,1j}+\frac{\sqrt{6}}{2}\mathcal{N}_{2i}\mathcal{C}^*_{L,2j}+{\sqrt{2}}\mathcal{N}_{3i}\mathcal{C}^*_{L,3j}.
\end{equation}
$|\vec{k}_\pi|$ is the 3-momentum norm of the pion in  $\chi^\pm_i$ rest frame given by
\begin{equation}
  |\vec{k}_\pi|=\frac{1}{2m_{\chi^\pm_i}} \Bigl[\Bigl(m^2_{\chi^\pm_i}-(m_{\chi^0_j}+m_\pi)^2\Bigr)\Bigl(m^2_{\chi^\pm_i}-(m_{\chi^0_j}-m_\pi)^2\Bigr)\Bigr]^{1/2}.
  \label{23}
\end{equation}

\begin{figure}[!t]
\centering
\subfigure[~Pure triplet case\label{disappearing track a}]
{\includegraphics[width=.48\textwidth,trim={0 15 0 10},clip]{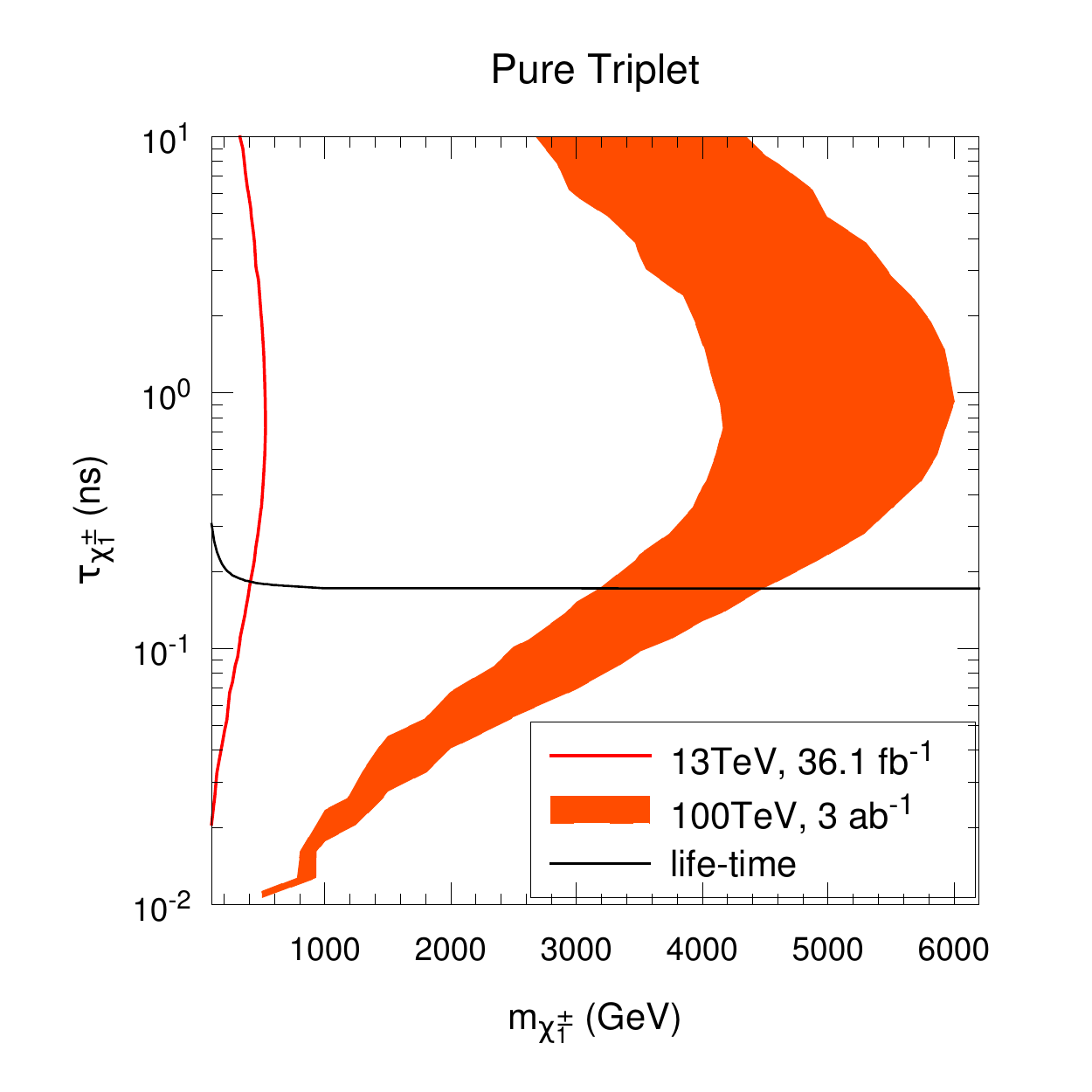}}
\subfigure[~Pure quadruplet case\label{disappearing track b}]
{\includegraphics[width=.48\textwidth,trim={0 15 0 10},clip]{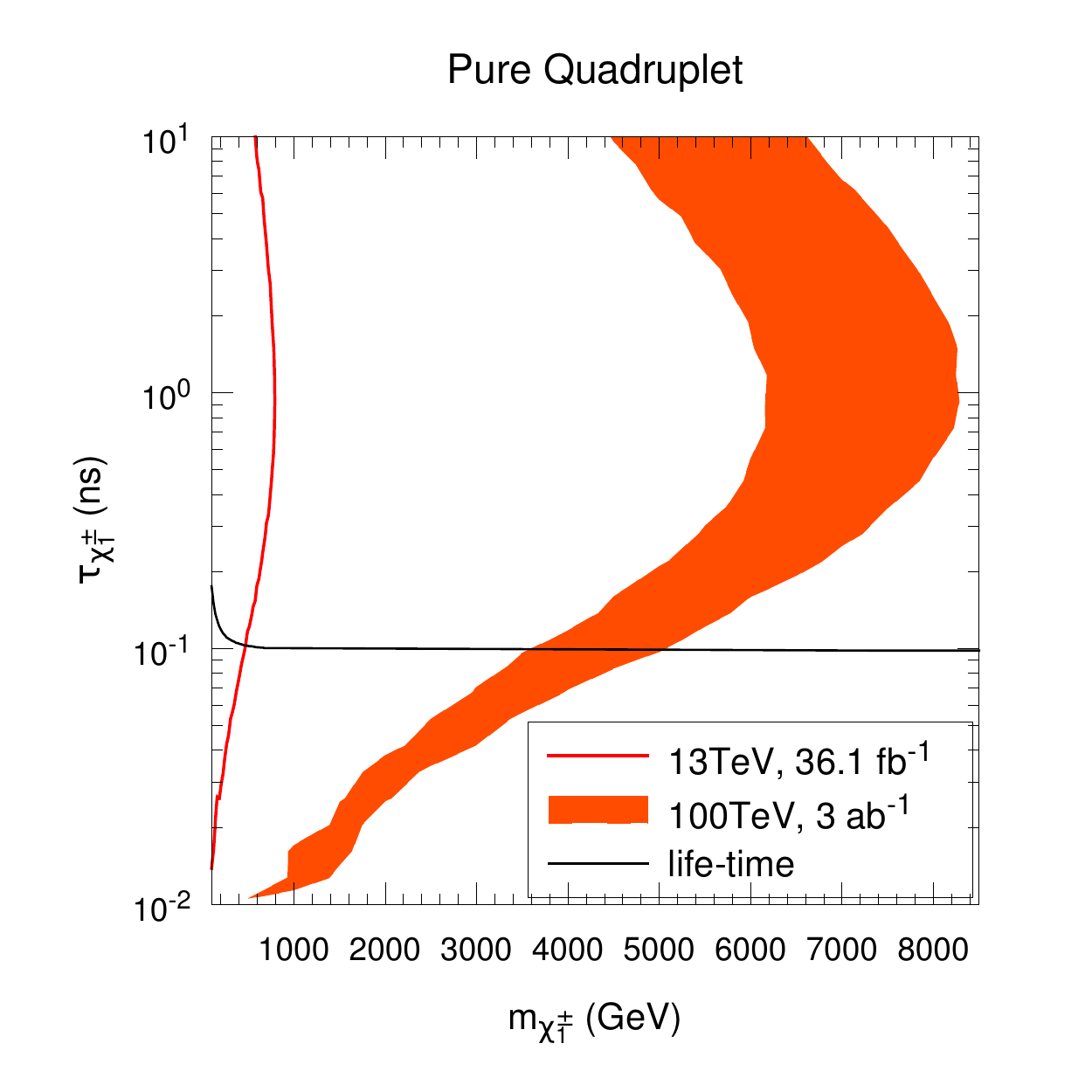}}
\caption{$95\%$ C.L. exclusion limits from the disappearing track channel in the $m_{\chi^\pm_1}$-$\tau_{\chi^\pm_1}$ plane for the pure triplet (a) and pure quadruplet (b) cases of the TQDM model.
The red solid lines represents the current exclusion limit from the ATLAS search based on a data set of $36.1~\si{fb}^{-1}$ at the 13~TeV LHC~\cite{ATLAS:2017bna}.
The red bands denote the ranges of the expected exclusion limit we estimate for the 100~TeV SPPC search with an integrated luminosity of $3~\si{fb^{-1}}$ by varying a background normalization factor in the range of 0.2-5.
The black solid lines indicate the predicted lifetime of $\chi^\pm_1$.}
\label{disappearing track}
\end{figure}

The predicted lifetime of $\chi^\pm_1$ as a function of $m_\mathrm{\chi^\pm_1}$ is indicated by the black solid lines in Figures \ref{disappearing track a} and \ref{disappearing track b} for the pure triplet and pure quadruplet cases, respectively.
In the plots, we also show the 95\% C.L. exclusion limit from the ATLAS search as the red lines.
When deriving these limit curves, the lifetime of $\chi^\pm_1$ for each  value of $m_{\chi^\pm_1}$ is treated as a free parameter.
Therefore, the intersections between the black and red lines provide lower limits on $m_{\chi^\pm_1}$.
For the pure triplet (quadruplet) case, $m_{\chi^\pm_1}\lesssim 410~(472)~\si{GeV}$ is excluded at $95\%$ C.L.
Such a constraint is much stronger than the current $\text{monojet}+\missET$ constraint, but it should be noted that the disappearing track channel is only sensitive to very restricted parameter regions where $m_{\chi^\pm_1}-m_{\chi^0_1}\lesssim 170~\si{MeV}$.

Below we discuss the prospect of the disappearing track channel at the 100~TeV SPPC.
The SM backgrounds are mainly contributed by the $W+\text{jets}$ and $t\bar{t}+\text{jets}$ processes.
Nonetheless, there are also three kinds of fake pixel tracklets that could contribute to the backgrounds~\cite{ATLAS:2017bna}:
\begin{enumerate}
\item A hadron undergoes a hard scattering with the inner detector and is not recognized as belonging to the same track.
\item A lepton emitting a hard photon could be identified as a disappearing tracklet.
\item A random combination of hits can be created by different nearby particles.
\end{enumerate}
All these known and other potentially unknown detector effects make us impossible to accurately simulate the backgrounds at the SPPC.
Instead, we can perform a simple estimation by rescaling the number of background events at the 13~TeV LHC~\cite{ATLAS:2017bna}, according to the event rates of the $W +\text{jets}$ and $t\bar{t}+\text{jets}$ backgrounds at the SPPC and LHC~\cite{Low:2014cba,Cirelli:2014dsa,Ostdiek:2015aga,Fukuda:2017jmk}.

\begin{table}[!t]
\setlength{\tabcolsep}{.5em}
\renewcommand{\arraystretch}{1.2}
\begin{tabular}{|c|c|c|c|c|}
\multicolumn{5}{c}{Pure triplet case, $m_Q=20~\si{TeV}$, $y_1=y_2=0$}\\
\hline
 & BMP-b1 & BMP-b2 & BMP-b3 & BMP-b4 \\
\hline
$m_T/\si{TeV}$ & 2.0 & 3.0 & 3.5 & 4.0 \\
\hline
\multicolumn{5}{c}{Pure quadruplet case, $m_T=200~\si{TeV}$, $y_1=y_2=0$}\\
\hline
 & BMP-c1 & BMP-c2 & BMP-c3 & BMP-c4 \\
\hline
$m_Q/\si{TeV}$ & 1.0 & 2.5 & 4.5 & 5.0 \\
\hline
\end{tabular}
\caption{Information of the signal BMPs in the disappearing track channel.}
\label{disappear_2}
\end{table}

\begin{figure}[!t]
\centering
\subfigure[~$\pT(j_1)$  distributions in the pure triplet case]
{\includegraphics[width=.48\textwidth,trim={0 10 0 15},clip]{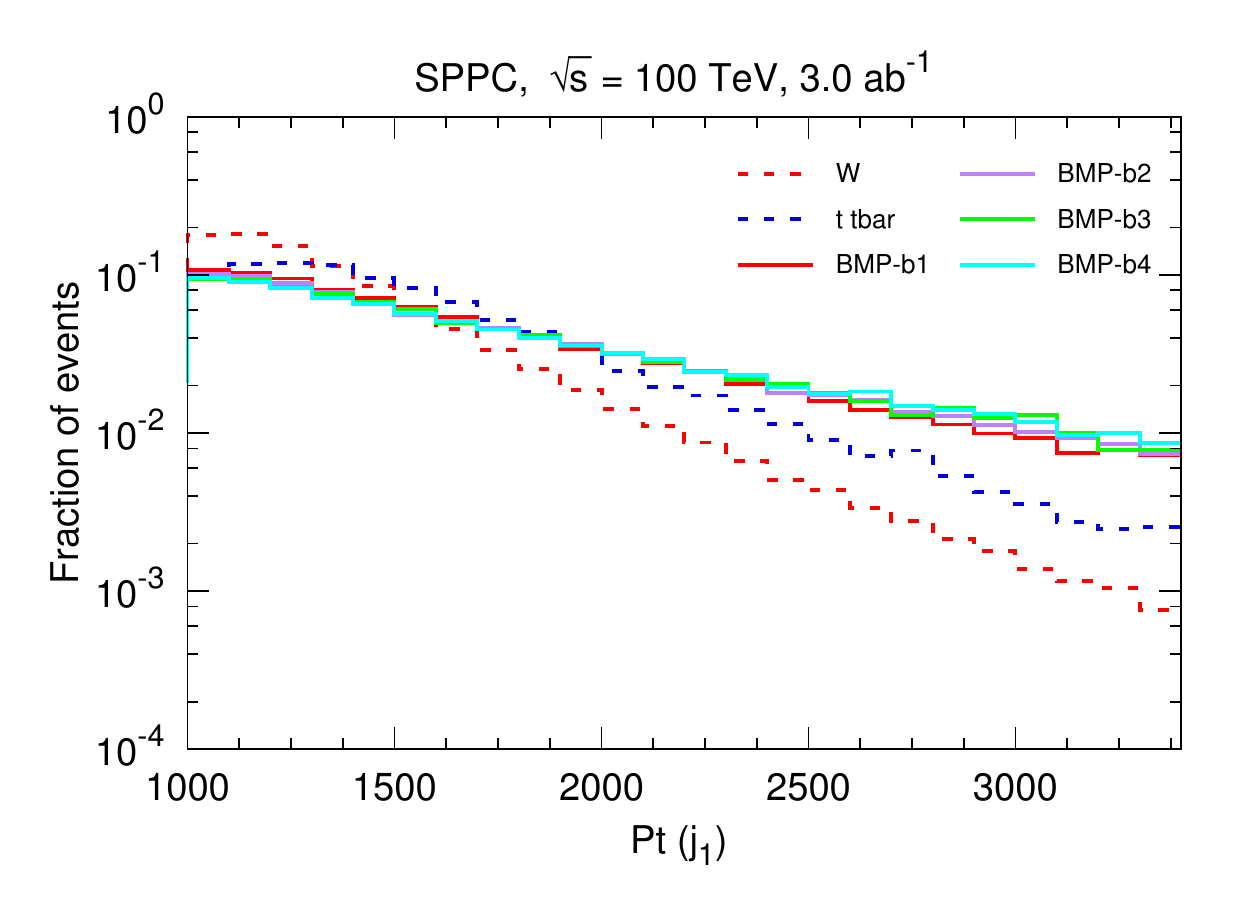}}
\subfigure[~$\missET$  distributions in the pure quadruplet case]
{\includegraphics[width=.48\textwidth,trim={0 10 0 15},clip]{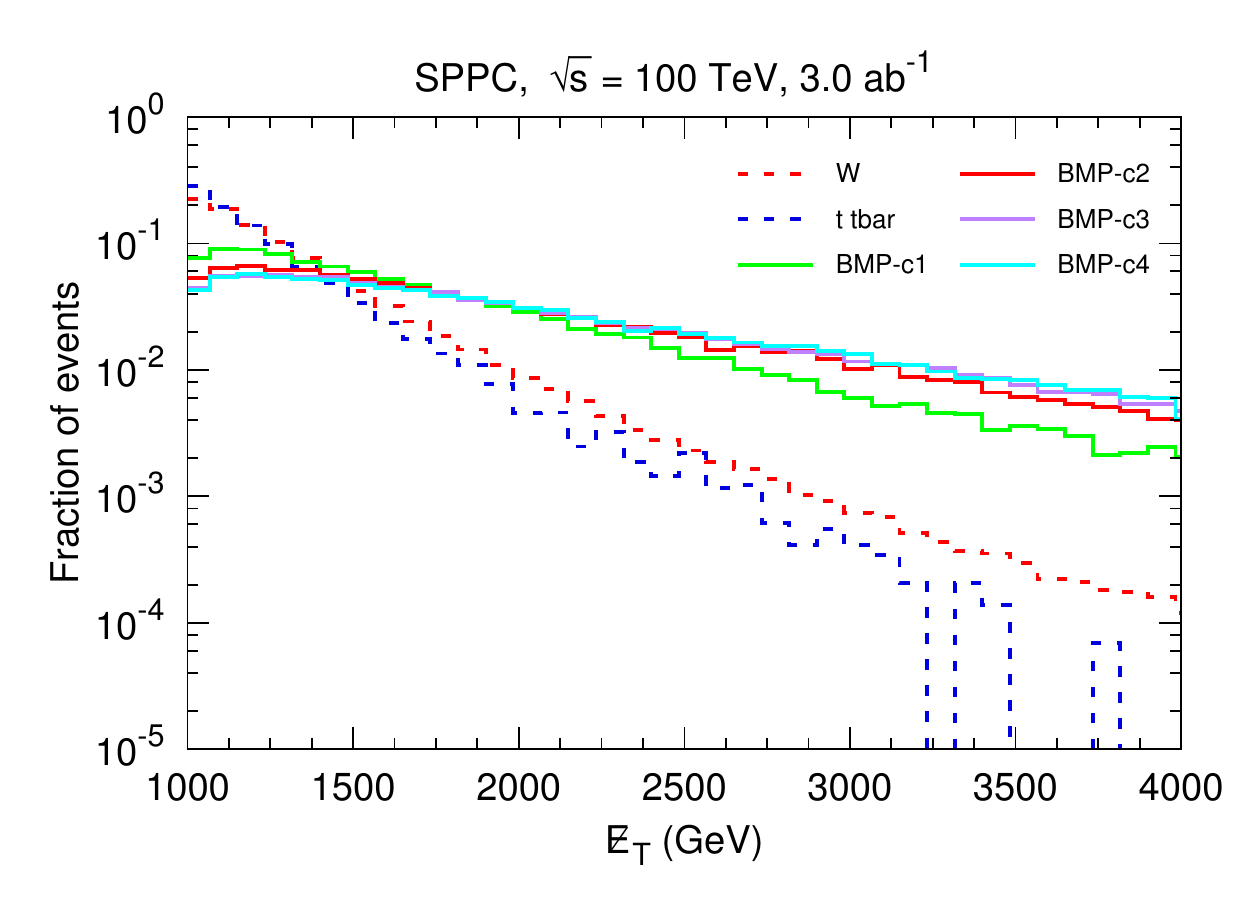}}
\caption{Normalized distributions of $\pT(j_1)$ in the pure triplet case (a) and $\slashed{E}_\mathrm{T}$ in the pure quadruplet case (b) at $\sqrt{s}=100~\si{TeV}$.
The dashed and solid lines represent the distributions for the backgrounds and signal BMPs, respectively.}
\label{disappearing track distribution}
\end{figure}

We choose several BMPs for the pure triplet and pure quadruplet cases, as listed in Table~\ref{disappear_2}.
Fig.~\ref{disappearing track distribution} shows the normalized distributions of $\pT(j_1)$ and $\slashed{E}_\mathrm{T}$ for the backgrounds and signal BMPs at $\sqrt{s}=100~\si{TeV}$.
According to these distributions, the cut thresholds for $\pT(j_1)$ and $\slashed{E}_\mathrm{T}$ can both be chosen to be $1500~\si{GeV}$.
We find that these thresholds are useful for both the pure triplet and quadruplet cases.
After applying all the cut conditions listed in the third column of Table~\ref{disappearing track cuts}, the expected background event number would be $\sim1$ with an integrated luminosity of $3~\si{ab^{-1}}$.  However, this number may be underestimated. In order to take into account the uncertainty in the background calculation, we adopt a range for the background estimation by rescaling this number by a factor of 0.2-5.

The expected 95\% C.L. exclusion limits at the SPPC are given as the red bands in Fig.~\ref{disappearing track}, according to the rescaling factor varying within 20\%-500\%.
In the pure triplet case, the SPPC search could reach up to $m_{\chi^\pm_1}\lesssim$ 3.2-4.5 TeV.
In the pure quadruplets, the lower limit of $m_{\chi^\pm_1}$ would be raised to 3.5-5.2 TeV.
Thus, the disappearing track channel at the SPPC has the potential to explore the whole range of $m_{\chi^0_1}$ allowed by the observed relic abundance for these two special cases.

\subsection{$\text{Multilepton}+\missET$ channel}

If the kinematics is allowed, charged and heavier neutral fermions in the dark sector would decay into leptonic states
via real or virtual $W$ and $Z$ bosons, leading to detectable signals in the $\text{multilepton}+\missET$ channel.
This channel has been widely used to explore SUSY models~\cite{ATLAS:2017uun,Aad:2014nua,CMS:2017fdz,Xiang:2016ndq}.
In this subsection, we just focus on the final state containing two
or three charged leptons associated with a large $\missET$.

In our simulation, the signal events come from $pp\to \chi\chi+\text{jets}$.
Main SM backgrounds arise from $WW+\text{jets}$, $WZ+\text{jets}$, $ZZ+\text{jets}$, and $t\bar{t}+\text{jets}$ production processes.
We have compared our MC results for SM backgrounds with the ATLAS MC results~\cite{ATLAS:2016uwq}, and find that they match with each other very well.

We used the latest ATLAS analysis~\cite{ATLAS:2017uun} as our primary reference for analyzing the $\text{multilepton}+\missET$ channel.
This channel can be categorized into three subchannels according to the numbers of leptons and jets in the final state, \textit{i.e.}, the $2\ell+\text{0jets}+\missET$, $2\ell+\text{jets}+\missET$, $3\ell+\missET$ channels.
There are some common reconstruction conditions for these channels:
\begin{itemize}
\item All jets must have $\pT > 20 ~\si{GeV}$ and $\vert\eta\vert < 2.8$.
\item Baseline electrons are required to have $\pT > 20 ~\si{GeV}$ and $\vert\eta\vert < 2.47$.
\item Baseline muons are required to have $\pT > 20 ~\si{GeV}$ and $\vert\eta\vert < 2.5$.
\item Central light-jets, which are tagged as $b$-jets, are required to have $\pT > 60 ~\si{GeV}$ and $\vert\eta\vert < 2.4$.
\item Central $b$-jets are required to have $\pT > 20 ~\si{GeV}$ and $\vert\eta\vert < 2.4$.
\end{itemize}
In the $2\ell+\text{0jets}+\missET$ channel, the leading and subleading leptons are required to have $\pT > 25 ~\si{GeV}$ and $\pT > 20 ~\si{GeV}$, respectively, and the signal events should not contain any central light-jet or central $b$-jet.
Other cut conditions for the signal regions in the three channels defined in the ATLAS analysis are listed in Tables~\ref{multilepton cut: a}, \ref{multilepton cut: b}, and\ref{multilepton cut: c}, where the numbers in brackets are the improved values we choose for the 100~TeV SPPC.

\begin{table}[!t]
\setlength{\tabcolsep}{.5em}
\renewcommand{\arraystretch}{1.2}
\begin{tabular}{|c|c|c|}
\multicolumn{3}{c}{$2\ell+\text{0jets}+\missET$ inclusive signal regions} \\
\hline
$m_\mathrm{T2}/\si{GeV}$ & $m_{\ell\ell}/\si{GeV}$ & Bin Order \\
\hline
$> 100$ & $> 111$ & 1 \\
$> 130$ & $> 300 ~(>200)$ & 2 \\
$> 100$ & - & 3 \\
$> 150$ & - & 4 \\
$> 200$ & - & 5 \\
$> 300$ & - & 6 \\
\hline
\end{tabular}
\caption{Cut conditions for the $2\ell+\text{0jets}+\missET$ inclusive signal regions. The number written in brackets is used for the 100~TeV SPPC, while other numbers are used in the ATLAS analysis at the 13~TeV LHC~\cite{ATLAS:2017uun}.}
\label{multilepton cut: a}
\end{table}

\begin{table}[!t]
\setlength{\tabcolsep}{.5em}
\renewcommand{\arraystretch}{1.2}
\begin{tabular}{|c|c|c|c|c|}
\multicolumn{5}{c}{$2\ell+\text{jets}+\missET$ signal regions} \\
\hline
Bin Order & 7 & 8 & 9* & 10* \\
\hline
$n_{\text{light-jets}}$ & $\ge 2$ & $\ge 2$ & 2 & 3-5 \\
$m_{\ell\ell}/\si{GeV}$ & 81-101 & 81-101 & 81-101 & 86-96 \\
$m_{jj}/\si{GeV}$ & 70-100 & 70-100 & 70-90 & 70-90 \\
$\slashed{E}_\mathrm{T}\ ~\si{GeV}$ & $>150$ & $>250$ & $>100$ & $>100$ \\
$\pT^{Z}/\si{GeV}$ & $>80~(>125)$ & $>80~(>125)$ & $>60$ & $>40$ \\
$\pT^{W}/\si{GeV}$ & $>100~(>130)$ & $>100~(>130)$ &  &  \\
$m_\mathrm{T2}/\si{GeV}$ & $>100$ & $>100$ &  &  \\
$\Delta R_{jj}$ & $<1.5$ & $<1.5$ &  & $<2.2$ \\
$\Delta R_{\ell\ell}$ & $<1.8$ &  &  &  \\
$\Delta \phi{(\slashed{\vec{p}}_\mathrm{T}, Z)}$ &  &  & $<0.8$ &  \\
$\Delta \phi{(\slashed{\vec{p}}_\mathrm{T}, W)}$ & 0.5-3.0 & 0.5-3.0 & $>1.5$ &  \\
$\slashed{E}_\mathrm{T}/p_\mathrm{T}^{Z}$ &  &  & 0.6-1.6 &  \\
$\slashed{E}_\mathrm{T}/p_\mathrm{T}^{W}$ &  &  & $<0.8$ &  \\
$\Delta \phi{(\slashed{\vec{p}}_\mathrm{T}, j_1)}$ &  &  & $>2.6$ &  \\
$\vert\eta(Z)\vert$ &  &  & $<1.6$ &  \\
$p_\mathrm{T}^{j_3}/\si{GeV}$ &  &  & $<1.6$ &  \\
\hline
\end{tabular}
\caption{Cut conditions for the $2\ell+\text{jets}+\missET$ inclusive signal regions.
$j_1$ and $j_3$ denote the jets with the highest and third highest $\pT$, respectively.
$Z$ and $W$ mean the $Z$ and $W$ bosons constructed by the two leptons ($\ell\ell$) and the two jets ($jj$), respectively
The numbers in brackets are used for the 100~TeV SPPC, while others are used in the ATLAS analysis at the 13~TeV LHC~\cite{ATLAS:2017uun}.}
\label{multilepton cut: b}
\end{table}

\begin{table}[!t]
\setlength{\tabcolsep}{.5em}
\renewcommand{\arraystretch}{1.2}
\begin{tabular}{|c|c|c|c|c|c|c|c|}
\multicolumn{8}{c}{$3\ell+\missET$ binned signal region} \\
\hline
$m_{\text{SFOS}}/\si{GeV}$ & $\slashed{E}_\mathrm{T}/\si{GeV}$ & $p_\mathrm{T}^{\ell_3}/\si{GeV}$ & $n_{\text{light-jets}}$ & $m_\mathrm{T}^{\text{min}}/\si{GeV}$ & $p_\mathrm{T}^{\ell\ell\ell}/\si{GeV}$ & $p_\mathrm{T}^{j_1}/\si{GeV}$ & bin order \\
\hline
81.2-101.2 & 60-120 & & 0 & $> 110$ &  &  & 11 \\
81.2-101.2 & 120-170 & & 0 & $> 110$ &  &  & 12 \\
81.2-101.2 & $>170$ & & 0 & $> 110$ &  &  & 13 \\
81.2-101.2 & 120-200 & & $\ge 1$ & $> 110$ & $<120$ & $>70$ & 14 \\
81.2-101.2 & $>200$ & & $\ge 1$ & 110-160 &  &  & 15 \\
81.2-101.2 & $>200$ & $>35$ & $\ge 1$ & $ >160$ &  &  & 16 \\
\hline
\end{tabular}
\caption{Cut conditions for the $3\ell+\missET$ binned signal regions in the ATLAS analysis at the 13~TeV LHC~\cite{ATLAS:2017uun}.
$\ell_3$ denotes the lepton with the third highest $\pT$,
while $j_1$ denotes the jet with the highest $\pT$.}
\label{multilepton cut: c}
\end{table}

Note that in the $2\ell+\text{0jets}+\missET$ channel, the kinematic variable $m_\mathrm{T2}$~\cite{Lester:1999tx,Barr:2003rg,Cheng:2008hk} is utilized instead of $\missET$ in order to effectively suppress backgrounds.
Thus, this channel is sensitive to signal processes like $\chi^+_2\chi^-_2 \to W^+(\to \ell^+ \nu)W^-(\to \ell^{'-} \bar\nu)\chi_1^0\chi_1^0$.
In the $2\ell+\text{jets}+\missET$ channel, two same-flavor opposite-sign (SFOS) leptons are used to reconstruct a $Z$ boson, while two jets are used to reconstruct a $W$ boson. The cut conditions utilize several kinematic variables related to the reconstructed $Z$ and $W$ bosons.
Such a channel is useful for searching signals like $\chi^0_2\chi^\pm_2 \to Z(\to \ell^+ \ell^-)W^\pm(\to jj)\chi_1^0\chi_1^0$.
In the $3\ell+\missET$ channel, a $Z$ boson is also reconstructed via two SFOS leptons, and the transverse mass $m_\mathrm{T}$, which should not be confused with the parameter $m_T$ in the TQDM model, is utilized for suppressing backgrounds.
This channel would be sensitive to signals like $\chi^0_2\chi^\pm_2 \to Z(\to \ell^+ \ell^-)W^\pm(\to \ell \nu)\chi_1^0\chi_1^0$.

\begin{figure}[!t]
\centering
\subfigure[~$y_1=y_2=0.5$\label{multilepton:a}]
{\includegraphics[width=.48\textwidth,trim={0 15 0 10},clip]{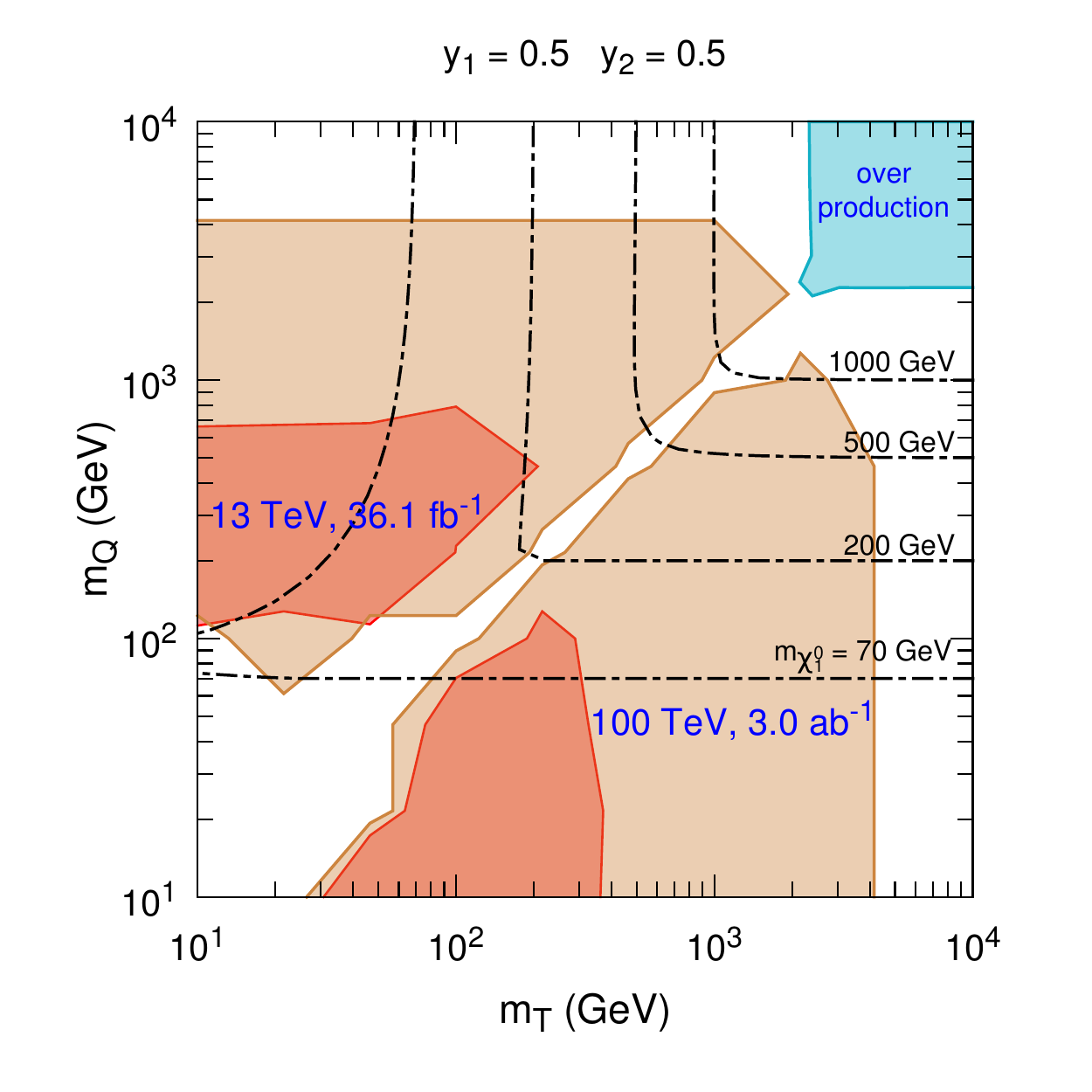}}
\subfigure[~$y_1=1.0$, $y_2=0.5$\label{multilepton:b}]
{\includegraphics[width=.48\textwidth,trim={0 15 0 10},clip]{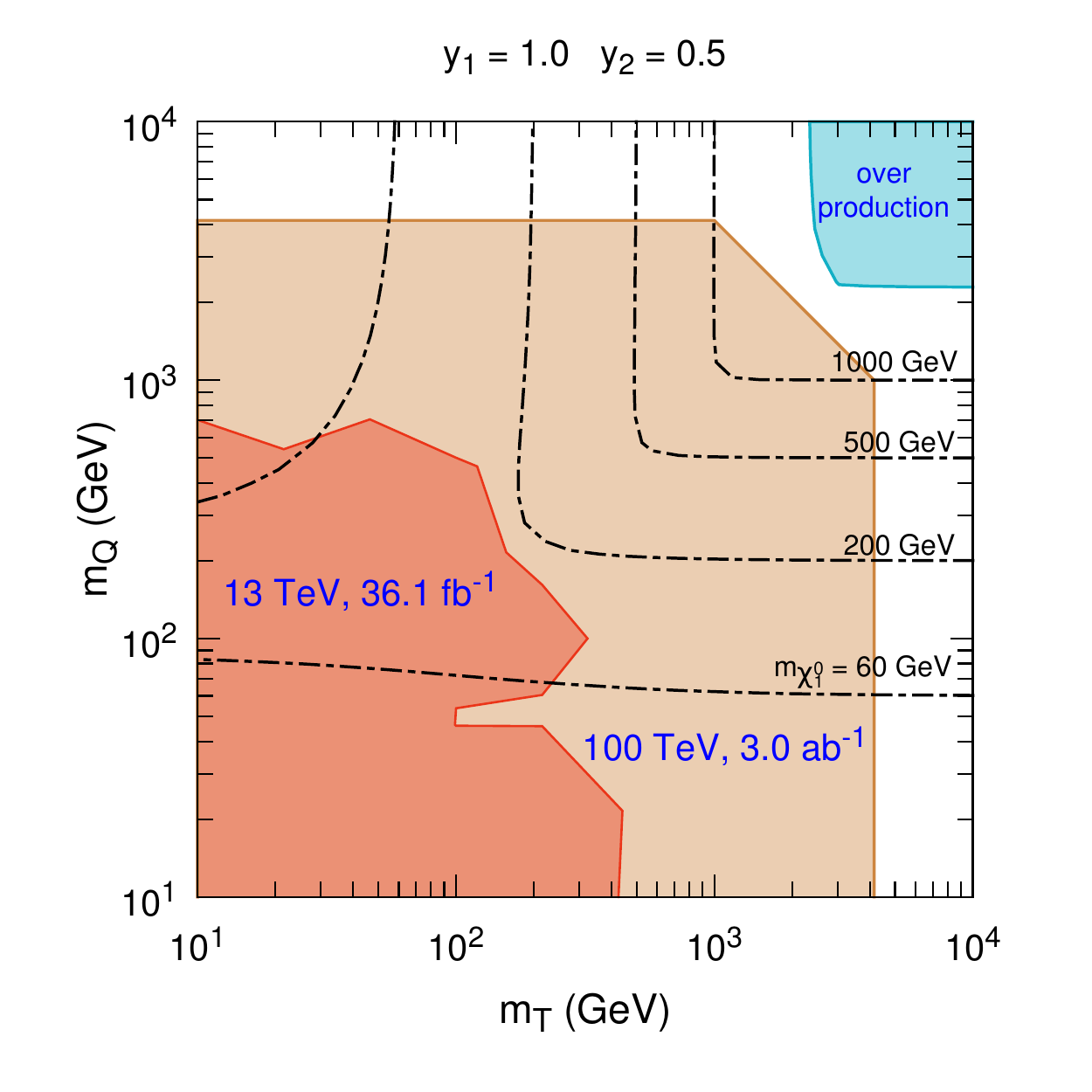}}
\caption{LHC Constraints and SPPC sensitivities from the $\text{multilepton}+\missET$ channel in the $m_T$-$m_Q$ plane with the fixed Yukawa couplings of $y_1=y_2=0.5$ (a) and $(y_1,y_2)=(1.0,0.5)$ (b).
The orange regions are excluded at 95$\%$ C.L. by the ATLAS search at the 13~TeV LHC with a data set of $36.1~\si{fb^{-1}}$~\cite{ATLAS:2017uun},
while the canary yellow regions are expected to be excluded at 95$\%$ C.L. by the 100~TeV SPPC with an integrated luminosity of $3~\si{ab^{-1}}$.
The blue regions are excluded because of overproduction of dark matter in the early Universe.}
\label{multilepton constrain}
\end{figure}

The constraint from the $\text{multilepton}+\missET$ channel on the TQDM model are shown in Fig.~\ref{multilepton constrain} for two Yukawa parameter sets of $y_1=y_2=0.5$ and $y_1=1$ and $y_2=0.5$.
The orange regions are excluded at 95\% C.L. by the ATLAS search at $\sqrt{s}=13~\si{TeV}$ with $36.1~\si{fb^{-1}}$ of data.
In the case of $y_1=y_2=0.5$, the custodial symmetry is respected, leading to a compressed particle spectrum in the regions with $m_T\sim m_Q$, small $m_T$, or small $m_Q$.
This means that the leptons from dark sector fermion decays would not be
energetic.
As a result, the $\text{multilepton}+\missET$ channel can hardly explore  these parameter regions.
In the case of $y_1=1.0$ and $y_2=0.5$, such parameter regions do not appear.

\begin{table}[!t]
\setlength{\tabcolsep}{.5em}
\renewcommand{\arraystretch}{1.2}
\begin{tabular}{|c|c|c|c|c|}
\hline
 & $y_1$ & $y_2$ & $m_T/\si{GeV}$ & $m_Q/\si{GeV}$ \\
\hline
BMP-d1 & 0.5 & 0.5 & 10.0  & 1000.0 \\
BMP-d2 & 0.5 & 0.5 & 215.4  & 464.2 \\
BMP-d3 & 0.5 & 0.5 & 464.2  & 100.0 \\
BMP-d4 & 0.5 & 0.5 & 464.2  & 1000.0 \\
\hline
\end{tabular}
\caption{Information of the signal BMPs in the $\text{multilepton} +\missET$ channel.}
\label{multilepton BMPs}
\end{table}

\begin{figure}[!t]
\centering
\subfigure[~$m_{\ell\ell}$ distributions for $2\ell+\text{0jets}+\missET$]
{\includegraphics[width=.48\textwidth,trim={0 10 0 10},clip]{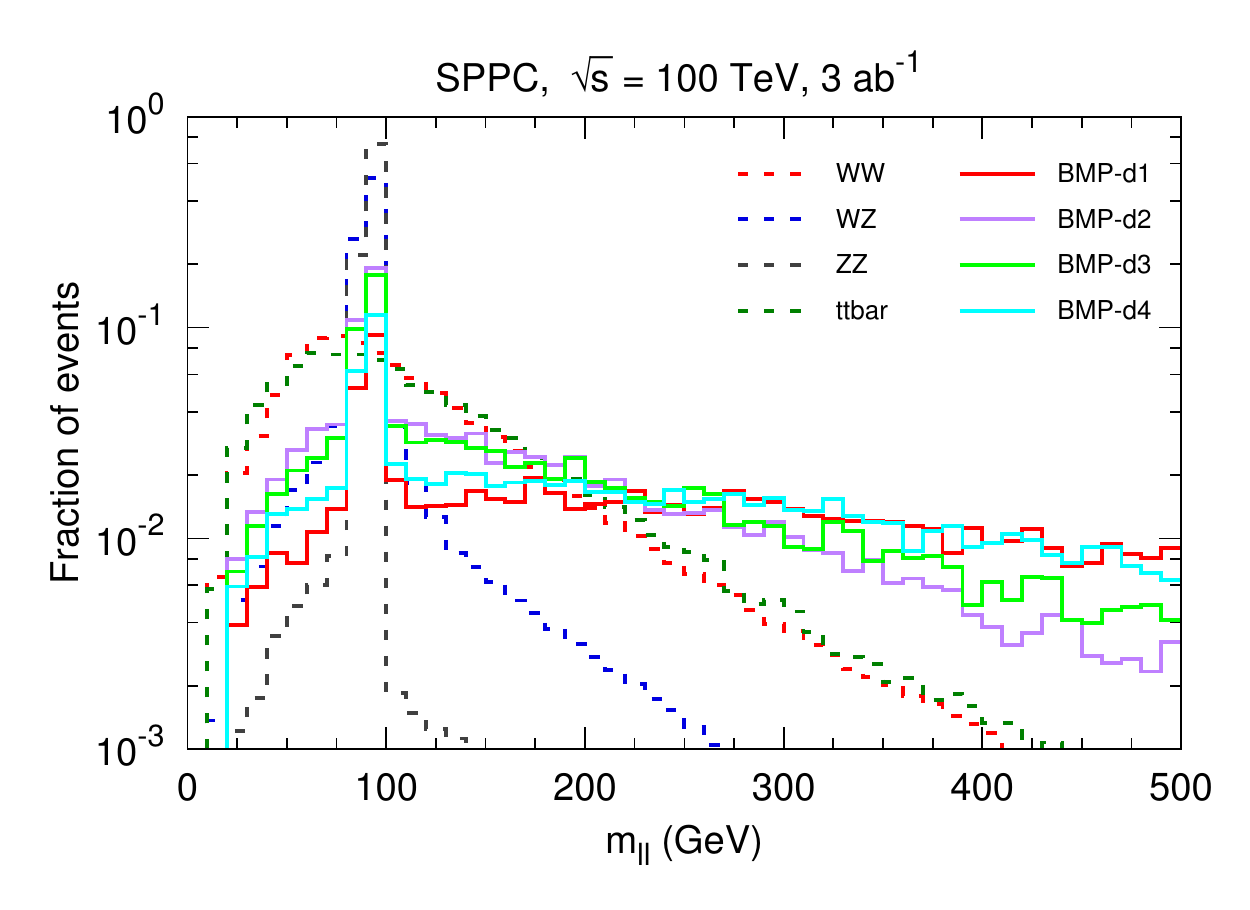}}
\subfigure[~$\pT^Z$ distributions for $2\ell+\text{jets}+\missET$]
{\includegraphics[width=.48\textwidth,trim={0 10 0 10},clip]{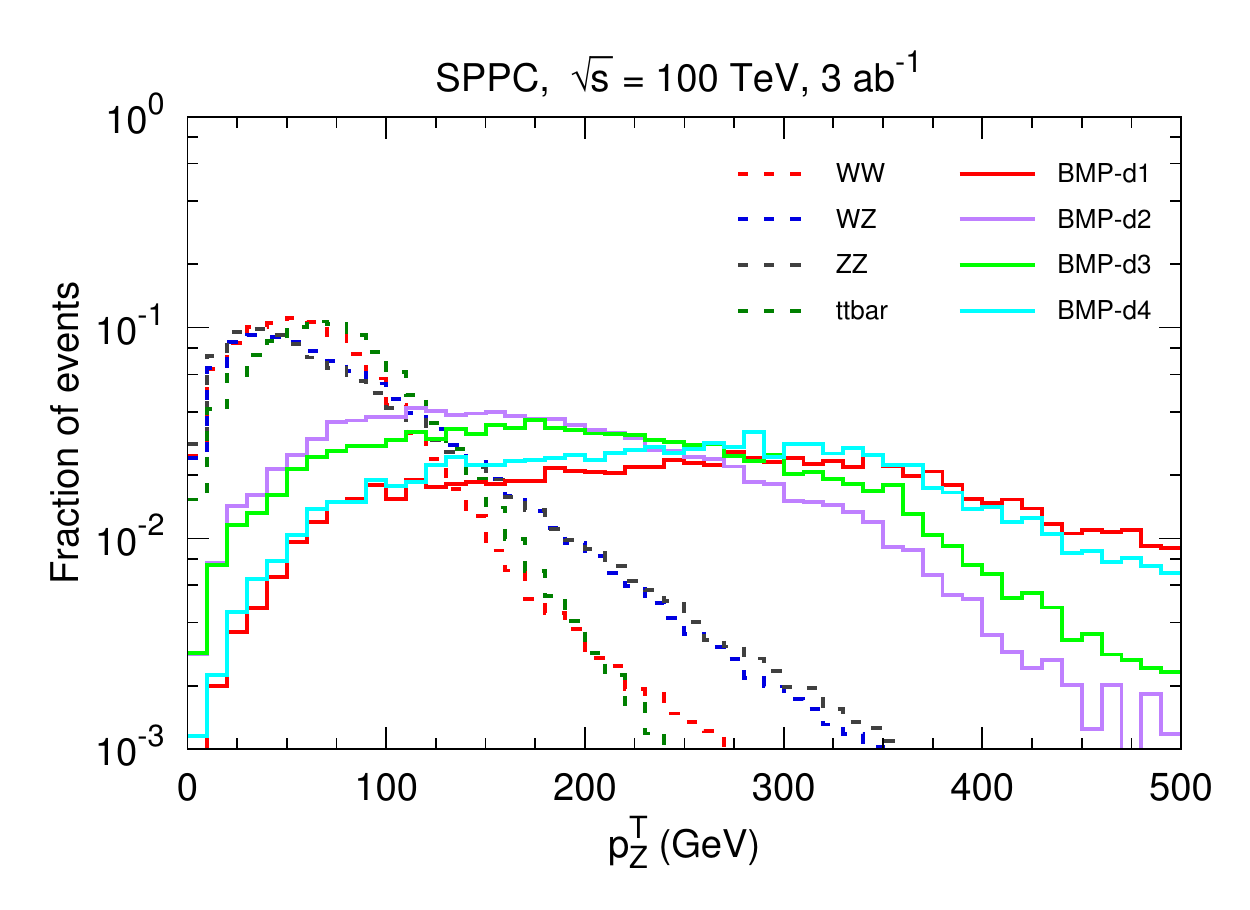}}
\caption{Normalized distributions of $m_{\ell\ell}$ in the $2\ell+\text{0jets}+\missET$ final state (a) and $\pT^Z$ in the $2\ell+\text{jets}+\missET$ final state (b) at $\sqrt{s}=100~\si{TeV}$. The dashed and solid lines represent the distributions for the backgrounds and signal BMPs, respectively.}
\label{multilepton distribution}
\end{figure}

Then we investigate the SPPC sensitivity. Four BMPs are adopted for studying cut thresholds, as listed in Table~\ref{multilepton BMPs}.
In Fig.~\ref{multilepton distribution}, distributions for the backgrounds and signal BMPs are presented.
Here we demonstrate the normalized distributions of two kinematic variables: the invariant mass of the lepton pair $m_{ll}$ for the $2\ell+\text{0jets}+\missET$ channel and the transverse momentum of the reconstructed $Z$ boson $\pT^Z$ for the $2\ell+\text{jets}+\missET$ channel.
Based on such distributions, we choose the cut thresholds of $m_{ll}$ and $\pT^Z$ to be 200 and $125~\si{GeV}$, respectively.
Other cut conditions can be found in Tables~\ref{multilepton cut: a} and \ref{multilepton cut: b}.

The canary yellow regions in Fig.~\ref{multilepton constrain} are expected to be excluded at $95\%$ C.L. by the $\text{multilepton}+\missET$ search at the 100~TeV SPPC with an integrated luminosity of $3~\si{ab^{-1}}$.
We find that such a search would probe the parameter regions up to $m_T \sim 4~\si{TeV}$ and $m_Q \sim 4~\si{TeV}$.
But this channel seems less powerful than the $\text{monojet}+\missET$ channel.

\section{CEPC Searches}
\label{sec:CEPC}

With a collision energy of $\sqrt{s} \sim$ 240-250 GeV and an integrated luminosity of $5~\si{ab^{-1}}$, more than one million Higgs bosons will be produced at the CEPC~\cite{CEPC-SPPCStudyGroup:2015csa}. The CEPC has a powerful capability to measure the detailed properties of the Higgs boson and explore BSM models through precision measurements.
In this section, we study loop effects on CEPC Higgs measurements induced by dark sector fermions in the TQDM model.

\subsection{$e^+ e^- \to Zh$ production}

The leading production process of the SM Higgs boson at the CEPC is the $Zh$ associated production $e^+ e^- \to Zh$.
The EW corrections to its cross section $\sigma_{Zh}$ at the next-to-leading order (NLO) in the SM can be found in \cite{Fleischer:1982af,Kniehl:1991hk,Denner:1992bc}.
Recently, the mixed QCD-EW $\mathcal{O}(\alpha\alpha_s)$ corrections to $\sigma_{Zh}$~\cite{Gong:2016jys,Sun:2016bel} and the ISR effects~\cite{Mo:2015mza} have also been studied.
For a data set of $5~\si{ab^{-1}}$, the relative precision of the $\sigma_{Zh}$ measurement would reach down to $\Delta \sigma_{Zh} / \sigma_{Zh} \sim 0.5\%$.

With such a high precision, new physics effects through loop corrections may manifest.
As the dark sector fermions in the TQDM model couple to both the Higgs boson and the EW gauge bosons, it is worth investigating their one-loop correction to $\sigma_{Zh}$.
We utilize the packages $\texttt{FeynArts~3.9}$~\cite{Hahn:2000kx}, $\texttt{FormCalc~9.4}$~\cite{Hahn:1998yk} and $\texttt{LoopTools~2.13}$~\cite{vanOldenborgh:1990yc} to calculate this correction at $\sqrt{s}=240~\si{GeV}$.
We adopt the on-shell renormalization scheme and neglect the mass and Yukawa coupling of the electron.
The deviation of $\sigma_{Zh}$ from the SM prediction can be expressed as
\begin{equation}
\Delta\sigma/\sigma_0 = |\sigma - \sigma_0|/\sigma_0.
\end{equation}
Here $\sigma_0=\sigma_\text{SM-LO} + \Delta \sigma_\text{SM-NLO}$ is the SM prediction including the leading-order cross section $\sigma_\text{SM-LO}$ and the NLO correction $\Delta \sigma_\text{SM-NLO}$.
$\sigma = \sigma_0 + \Delta \sigma_\text{TQ-NLO}$ involves the NLO contribution from the TQDM model, $\Delta \sigma_{\text{TQ-NLO}}$.
If the predicted $\Delta\sigma/\sigma_0$ is larger than $0.5\%$, the CEPC measurement should be able to probe such an effect of the TQDM model.

\begin{figure}[!t]
\centering
\subfigure[~$y_1=y_2=0.5$ \label{eehz:a}]
{\includegraphics[width=.48\textwidth,trim={0 25 0 25},clip]{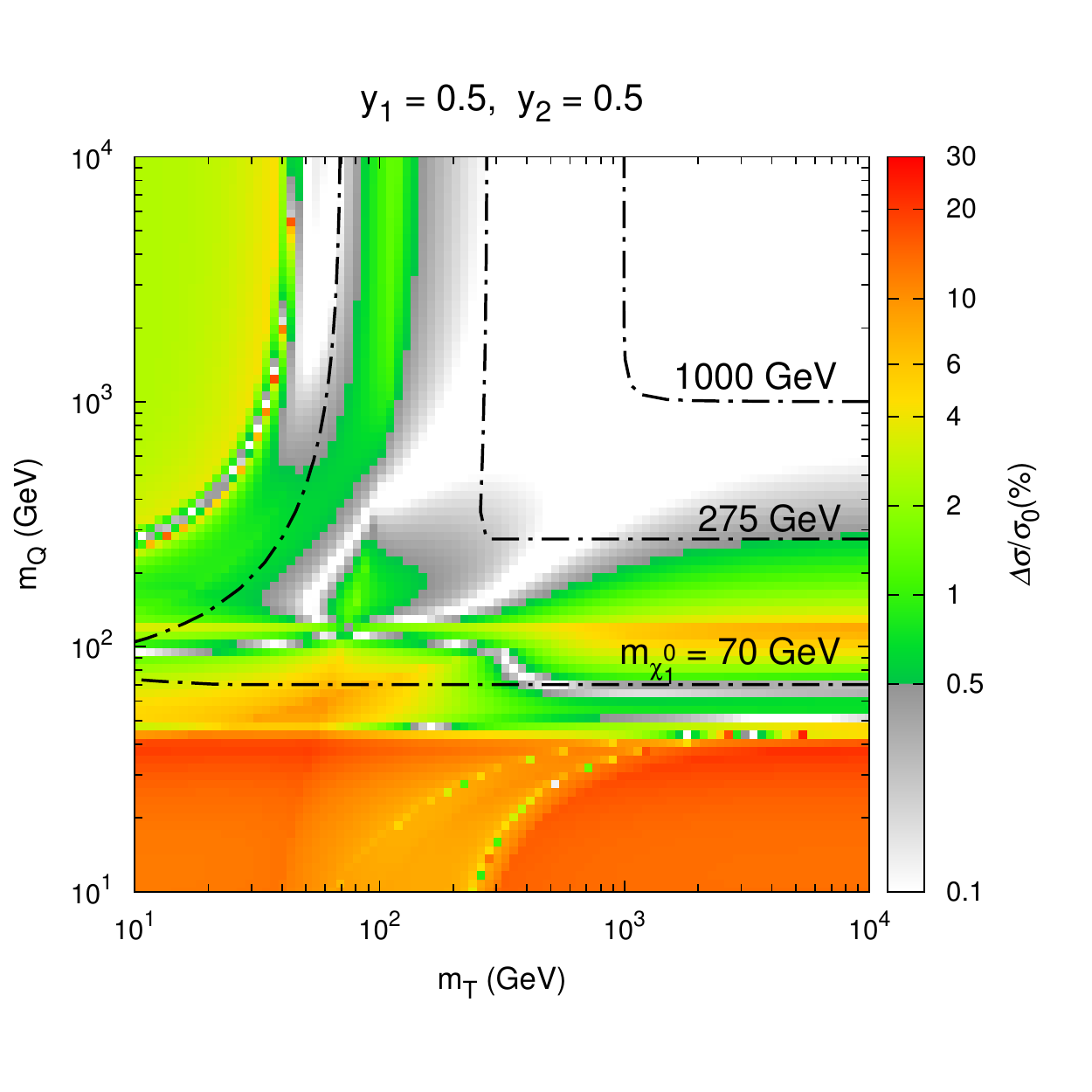}}
\subfigure[~$y_1=1.0$, $y_2=0.5$ \label{eehz:b}]
{\includegraphics[width=.48\textwidth,trim={0 25 0 25},clip]{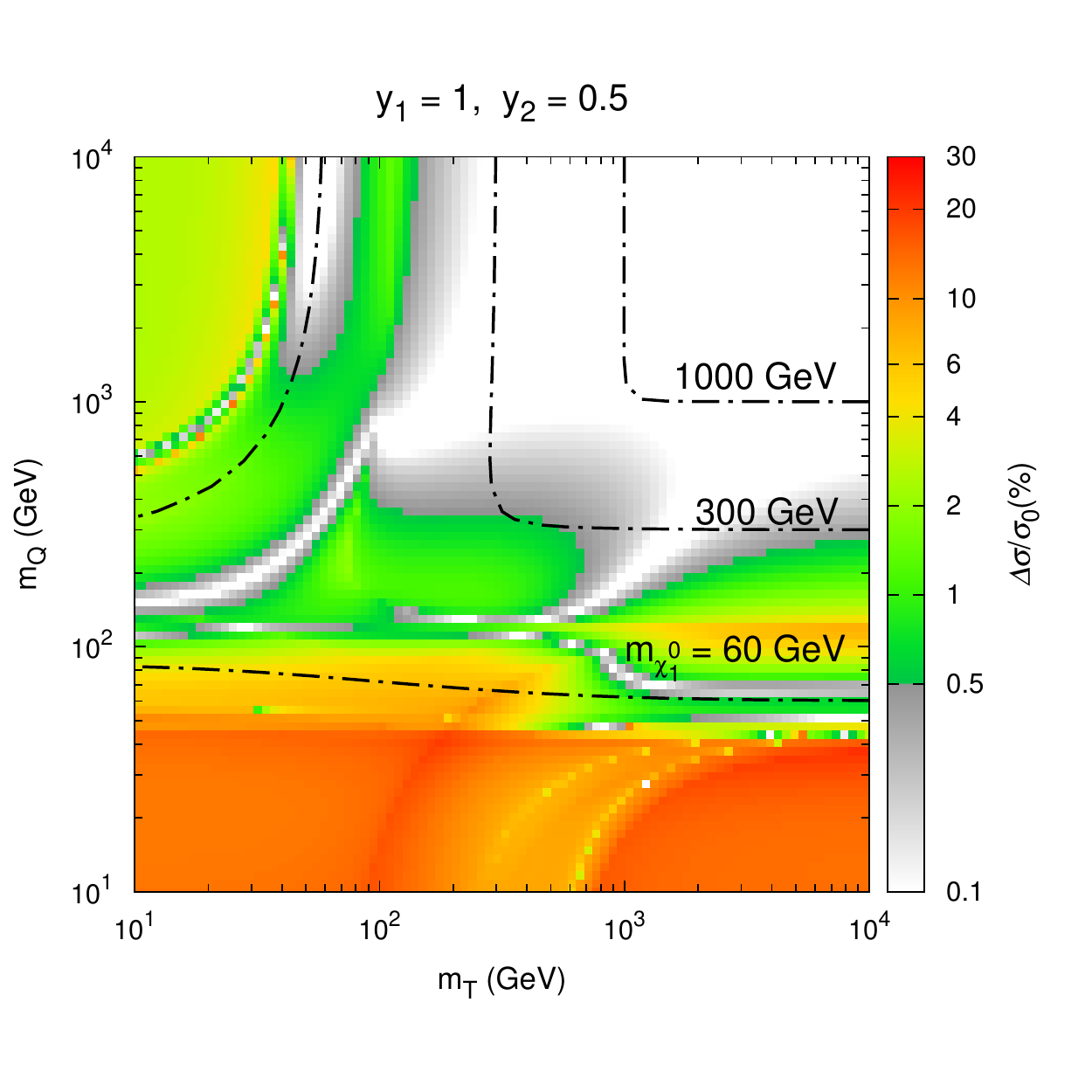}}
\subfigure[~$m_T=100~\si{GeV}$, $m_Q=400~\si{GeV}$ \label{eehz:c}]
{\includegraphics[width=.48\textwidth,trim={0 20 0 25},clip]{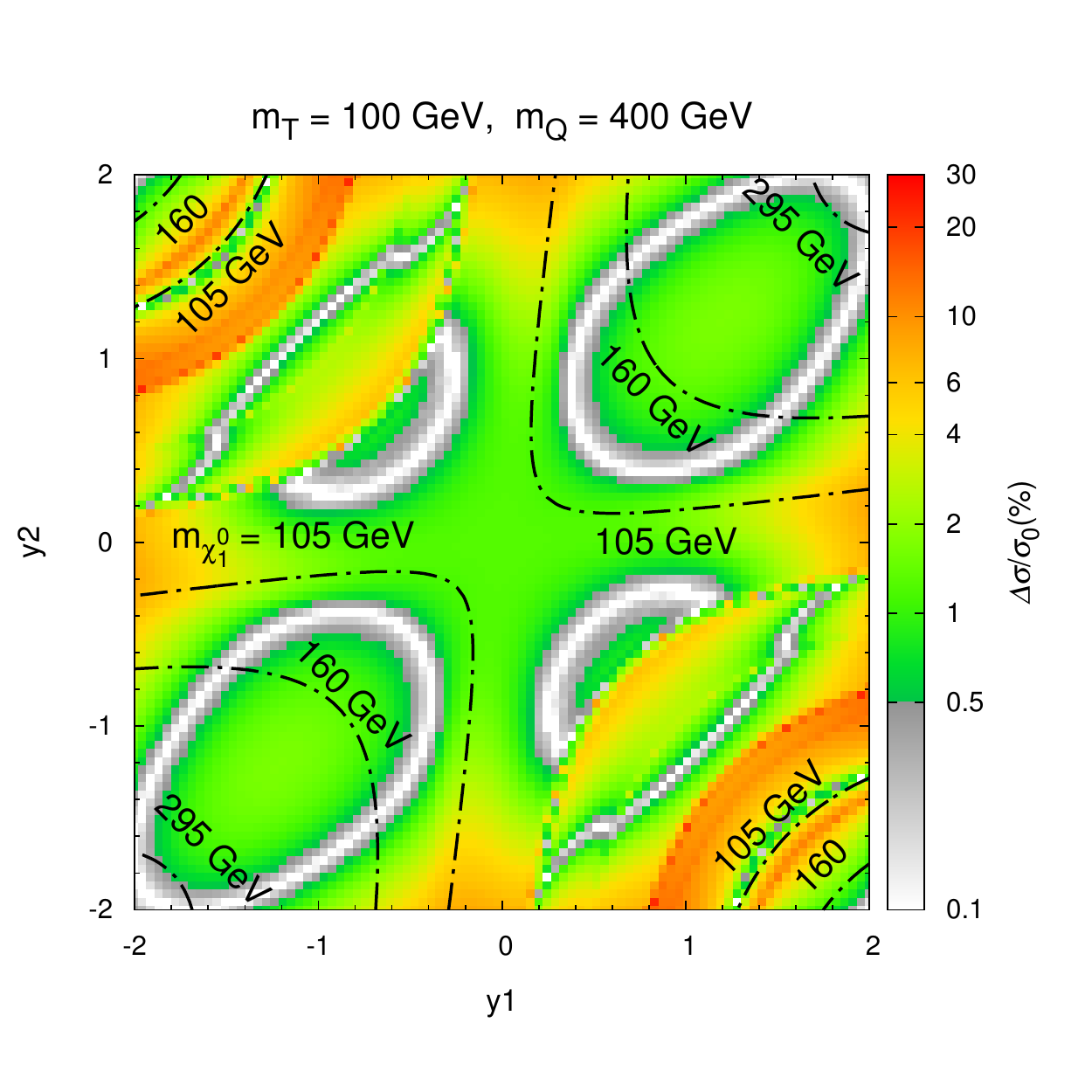}}
\subfigure[~$m_T=400~\si{GeV}$, $m_Q=200~\si{GeV}$ \label{eehz:d}]
{\includegraphics[width=.48\textwidth,trim={0 20 0 25},clip]{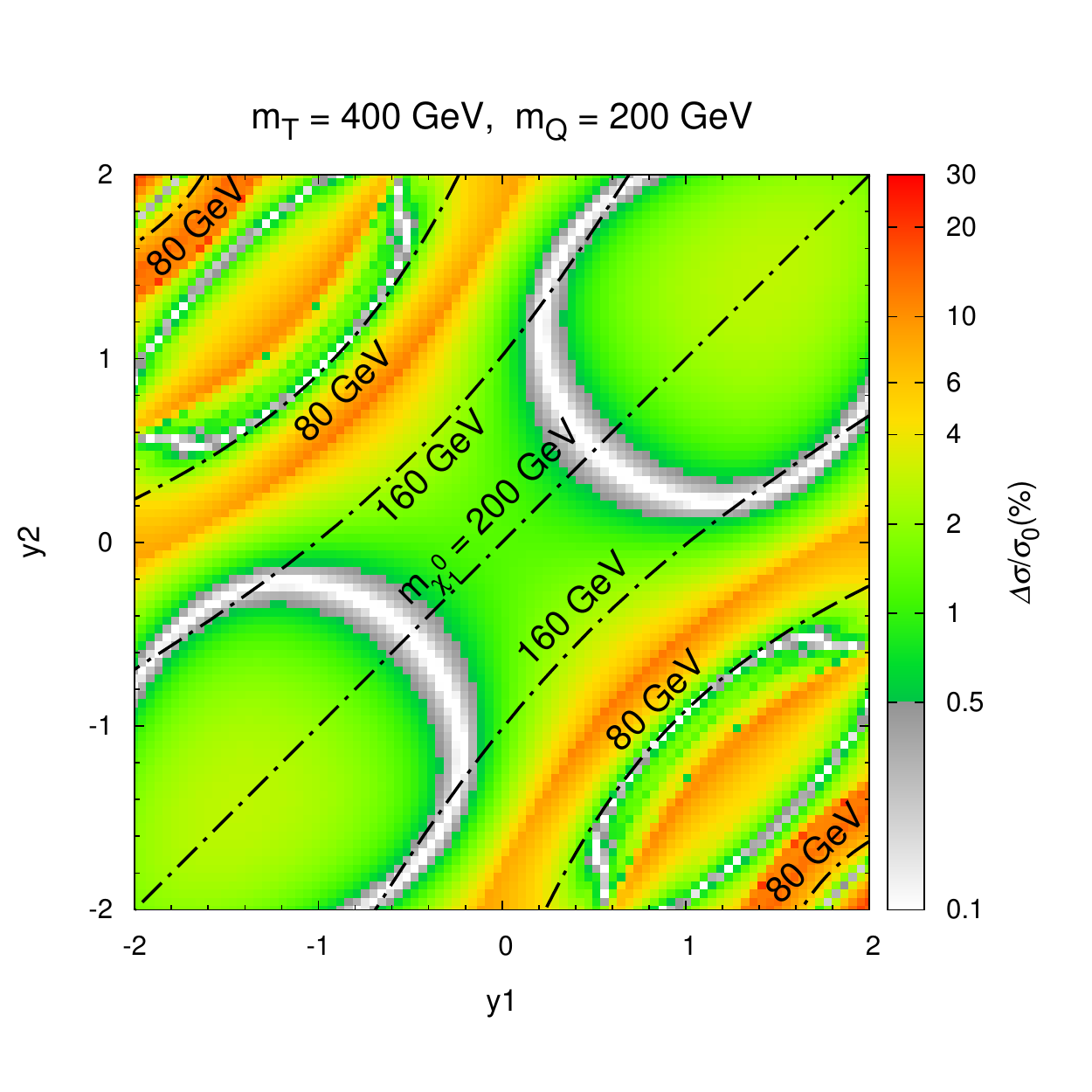}}
\caption{Heat maps for the relative deviation of the $e^+e^-\to Zh$ cross section $\Delta \sigma/\sigma_0$ predicted in the TQDM model.
The top and bottom panels show the results in the $m_T$-$m_Q$ and $y_1$-$y_2$ planes with the other two parameters fixed, respectively.
Colored and gray regions correspond to $\Delta \sigma/\sigma_0 > 0.5\%$ and $< 0.5\%$, respectively.
Dot-dashed lines denote the contour of $m_{\chi_1^0}$.}
\label{fig:eehz}
\end{figure}

We calculate $\Delta\sigma /\sigma_0$ for four benchmark cases, as shown in Fig.~\ref{fig:eehz}. For each case, we fix two parameters in $(m_T, m_Q, y_1, y_2)$ and vary the other two. The colored regions corresponding to $\Delta \sigma / \sigma_0 > 0.5\%$ may be
explored by the CEPC measurement, while the gray regions are beyond its capability.
The dot-dashed lines represent the mass of $\chi^0_1$.

In Figures \ref{eehz:a} and \ref{eehz:b}, the Yukawa couplings are fixed to be $y_1=y_2=0.5$ and $(y_1,y_2)=(1.0,0.5)$, respectively.
We can see that the CEPC would explore up to $m_{\chi^0_1}\sim$ 275-300 GeV for these two cases.
When $m_T\gtrsim 100~\si{GeV}$ and $m_Q\gtrsim 300~\si{GeV}$, all dark sector fermions become heavy, suppressing the Higgs effective interactions with the photon and $Z$.
Thus, the corrections in these region would not be significant.

It is promising to search the loop effects on $\sigma_{zh}$ for small $m_T$ and $m_Q$.
Thus, we fix $(m_T, m_Q)$ to be $(100, 400)~\si{GeV}$ and $(400, 200)~\si{GeV}$ in Figures \ref{eehz:c} and \ref{eehz:d}, respectively, to examine how $\Delta \sigma / \sigma_0$ varies with $y_1$ and $y_2$.
In these two cases, $\chi_1^0$ is dominated by either the triplet or the quadruplets.
We can see that $\Delta \sigma / \sigma_0 > 0.5\%$ almost holds in the whole $y_1$-$y_2$ plane.

\begin{figure}[!t]
\centering
\subfigure[~$y_1=0.5,~y_2=0.5$]
{\includegraphics[width=0.45\textwidth,trim={0 10 0 10},clip]{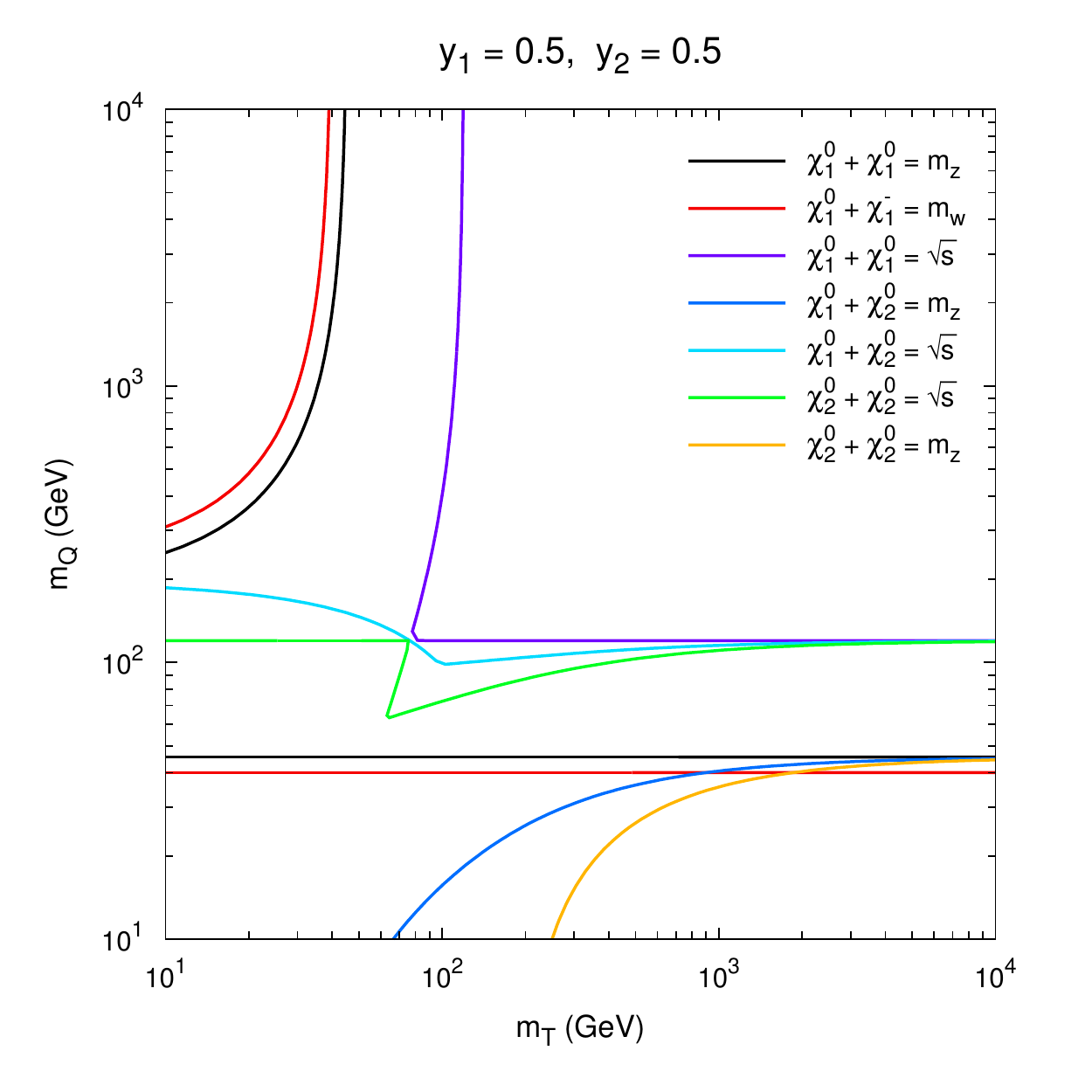}}
\subfigure[~$m_T=100~\si{GeV},~m_Q=400~\si{GeV}$]
{\includegraphics[width=0.435\textwidth]{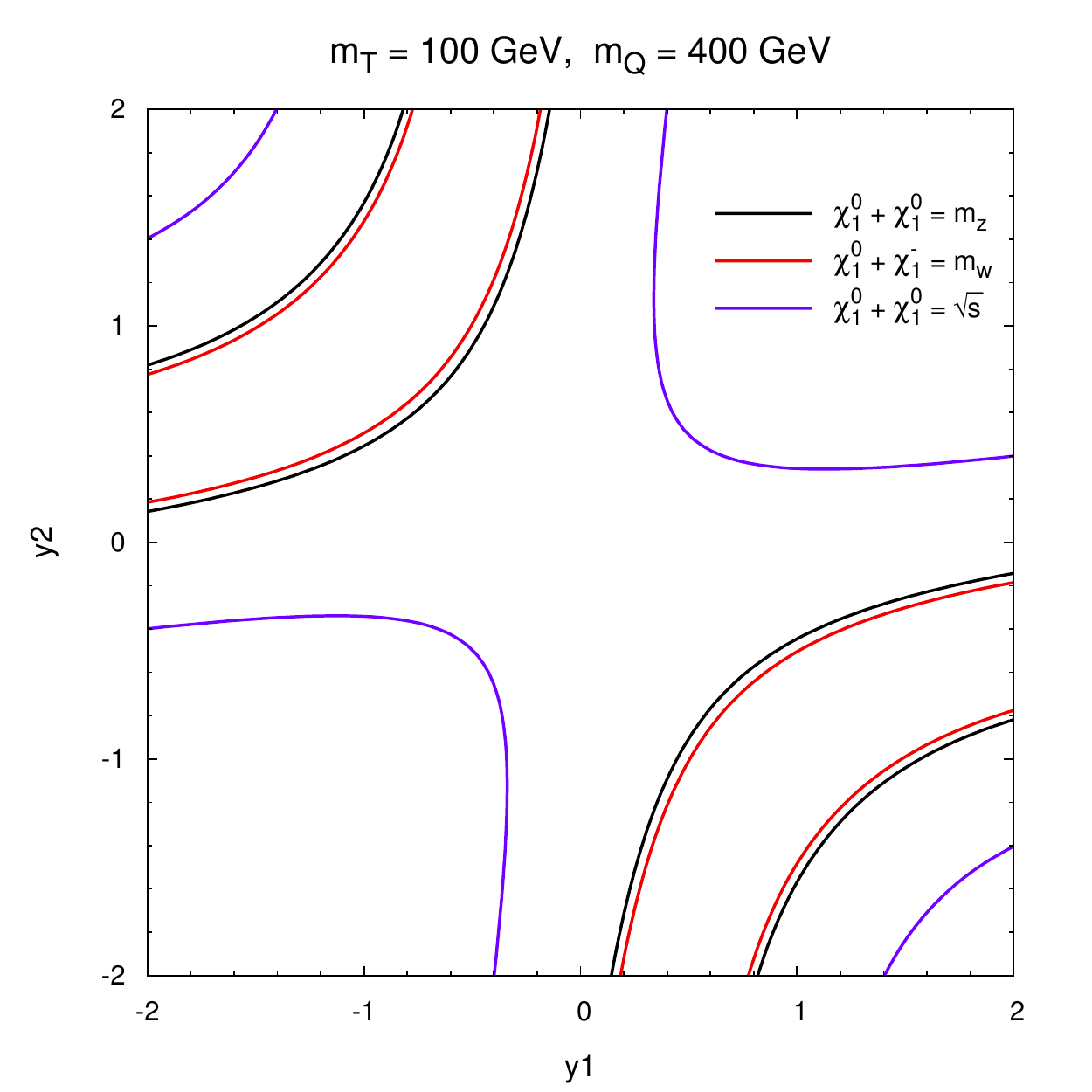}}
\caption{Contours for threshold conditions. The left and right panels correspond to the parameter sets in Fig.~\ref{eehz:a} and Fig.~\ref{eehz:c}, respectively.}
\label{eehz:exp}
\end{figure}

Note that the variation of $\Delta\sigma /\sigma_0$ in all these plots looks quite complicated.
This is mainly due to the threshold effect.
For instance, in the regions with $m_{\chi^0_i}+m_{\chi^0_j}\sim m_Z$, $m_{\chi^0_i}+m_{\chi^-_j}\sim m_W$, $m_{\chi^0_i}+m_{\chi^-_j}(m_{\chi^0_j})\sim \sqrt{s}$ and so on, the propagators of dark sector fermions meet some poles, and their contributions change dramatically.
In order to more explicitly analyze such a effect, we plot the contours indicating some threshold conditions in Fig.~\ref{eehz:exp}.
Comparing Figures \ref{eehz:a} and \ref{eehz:c} with Fig.~\ref{eehz:exp}, we can find that most of the $\Delta\sigma /\sigma_0$ structures match these contours.

\subsection{$h\to\gamma\gamma$ decay}

In the SM, the Higgs boson decay into diphoton is induced by loops of the $W$ boson and heavy charged fermions.
The deviation of the $h\to\gamma\gamma$ partial width $\Gamma_{\gamma\gamma}$ from the SM prediction $\Gamma_{\gamma\gamma}^\mathrm{SM}$ can be characterized by a ratio $\kappa_\gamma \equiv \Gamma_{\gamma\gamma}/ \Gamma_{\gamma\gamma}^\mathrm{SM}$.
The CEPC can measure this ratio with a high precision of $4.7\%$ for an integrated luminosity of $5~\si{ab^{-1}}$~\cite{CEPC-SPPCStudyGroup:2015csa}.
In the TQDM model, the singly charged fermions $\chi^\pm_i$ couple to both the Higgs boson and the photon, and hence modify the value of $\kappa_\gamma$ from 1.
Therefore, the measurement of $h\to\gamma\gamma$ could be useful for exploring the TQDM model.

Note that the doubly charged fermion $\chi^{\pm\pm}$, which is only constructed by the quadruplets, does not contribute to $\Gamma_{\gamma\gamma}$. This is because an $\SUtwoL$ invariant Yukawa term $\sim \mathrm{QTH}$ must be built from a triplet and a quadruplet.
The $h\to\gamma\gamma$ partial decay width can be cast into the form of \cite{Ellis:1975ap,Shifman:1979eb,Djouadi:2005gi,Xiang:2017yfs}
\begin{equation}
\Gamma_{\gamma\gamma}=\frac{G_\mathrm{F} \alpha^2 m^3_h}{128 \sqrt{2}\pi^3}\biggl| \sum_{f} N_c Q^2_f A_{1/2}(\tau_f)+A_1(\tau_W)+ \sum_{i} \frac{G_{h,ii}v}{m_{\chi^\pm_i}}A_{1/2}(\tau_{\chi^\pm_i})\biggr|^2,
  \label{20}
\end{equation}
where $\alpha$ is the fine-structure constant, $N_c$ is the color factor, $Q_f$ is the electric charge of an SM fermion $f$, and $G_{h,ii}$ is the $\chi^-_i \chi^+_i h$ coupling, which can be read from Eq.~\eqref{higgs}:
\begin{equation}
  G_{h,ii}=\text{Re}\left(-\frac{y_1}{\sqrt{2}}\mathcal{C}_{\text{L},2i}\mathcal{C}_{\text{R},1i}+\frac{y_1}{\sqrt{6}}\mathcal{C}_{\text{L},1i}\mathcal{C}_{\text{R},2i}+\frac{y_2}{\sqrt{6}}\mathcal{C}_{\text{L},3i}\mathcal{C}_{\text{R},1i}-\frac{y_2}{\sqrt{2}}\mathcal{C}_{\text{L},1i}\mathcal{C}_{\text{R},3i}\right).
  \label{34}
\end{equation}
The first two terms between the vertical bars are SM contributions, while the third term comes from the TQDM model.
$A_{1/2}$ and $A_1$ are the form factors for spin-1/2 and spin-1 particles:
\begin{equation}
    A_{1/2}(\tau)=2[\tau +(\tau - 1)f(\tau)]\tau^{-2} ,\quad
    A_1(\tau)=-[2\tau^2+3\tau+3(2\tau-1)f(\tau)]\tau^{-2},
  \label{16}
\end{equation}
where the function $f(\tau)$ and the parameters $\tau$ are defined as
\begin{eqnarray}
  f(\tau)&=&\begin{dcases}
    \text{arcsin}^2\sqrt{\tau}, & \tau \le 1,\\
    -\frac14 \biggr[\text{log}\frac{1+\sqrt{1-\tau^{-1}}}{1-\sqrt{1-\tau^{-1}}}-i\pi\biggr]^2, & \tau > 1,
  \end{dcases}
  \label{21}
\\
  \tau_W&\equiv&\frac{m^2_h}{4m^2_W},\quad \tau_f\equiv\frac{m^2_h}{4m^2_f},\quad \tau_{\chi^\pm_i}\equiv\frac{m^2_h}{4m^2_{\chi^\pm_i}}.
  \label{22}
\end{eqnarray}

\begin{figure}[!t]
\centering
\subfigure[~$y_1=y_2=0.5$]
{\includegraphics[width=0.43\textwidth,trim={0 10 0 10},clip]{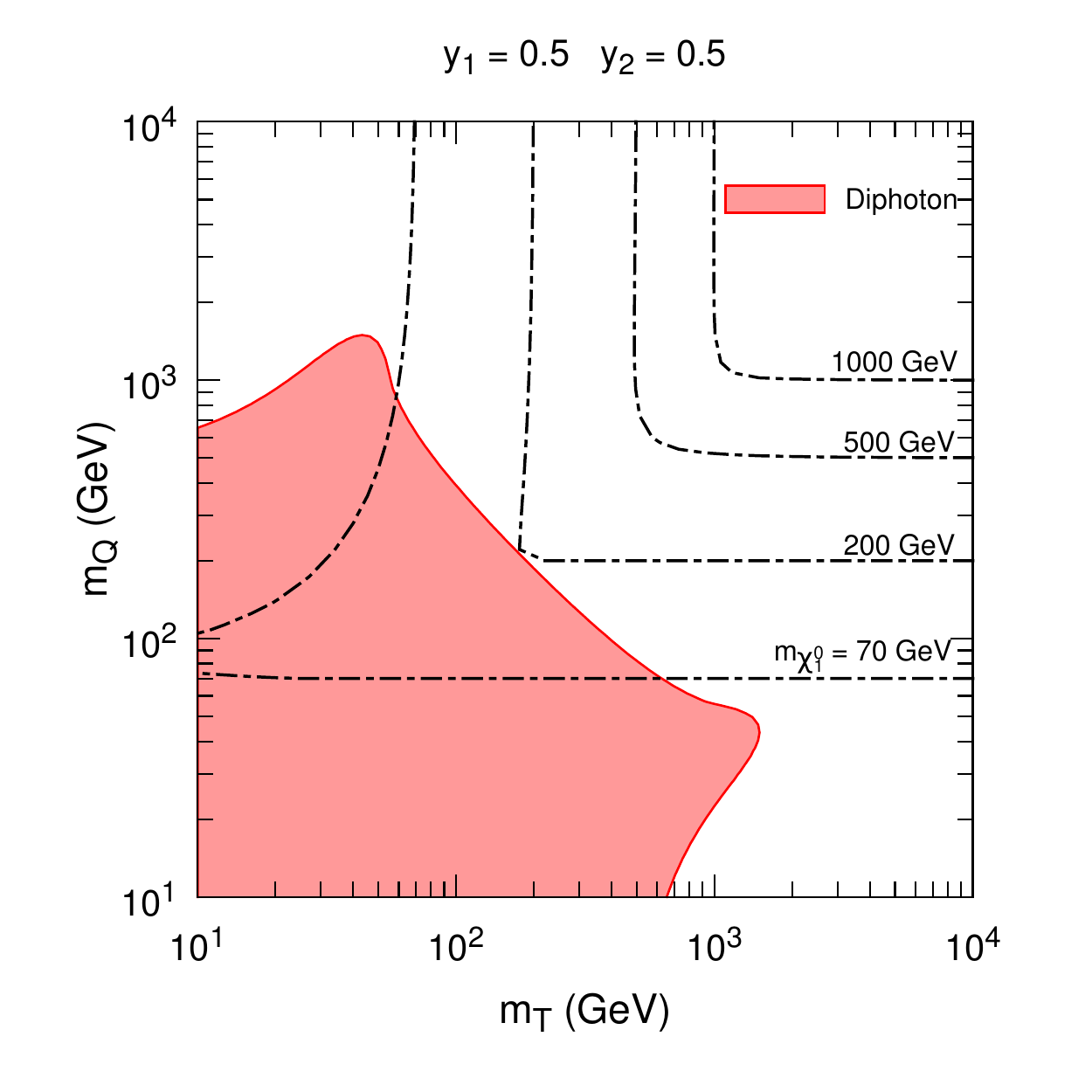}}
\hspace{0.8cm}
\subfigure[~$y_1=1.0$, $y_2=0.5$]
{\includegraphics[width=0.43\textwidth,trim={0 10 0 10},clip]{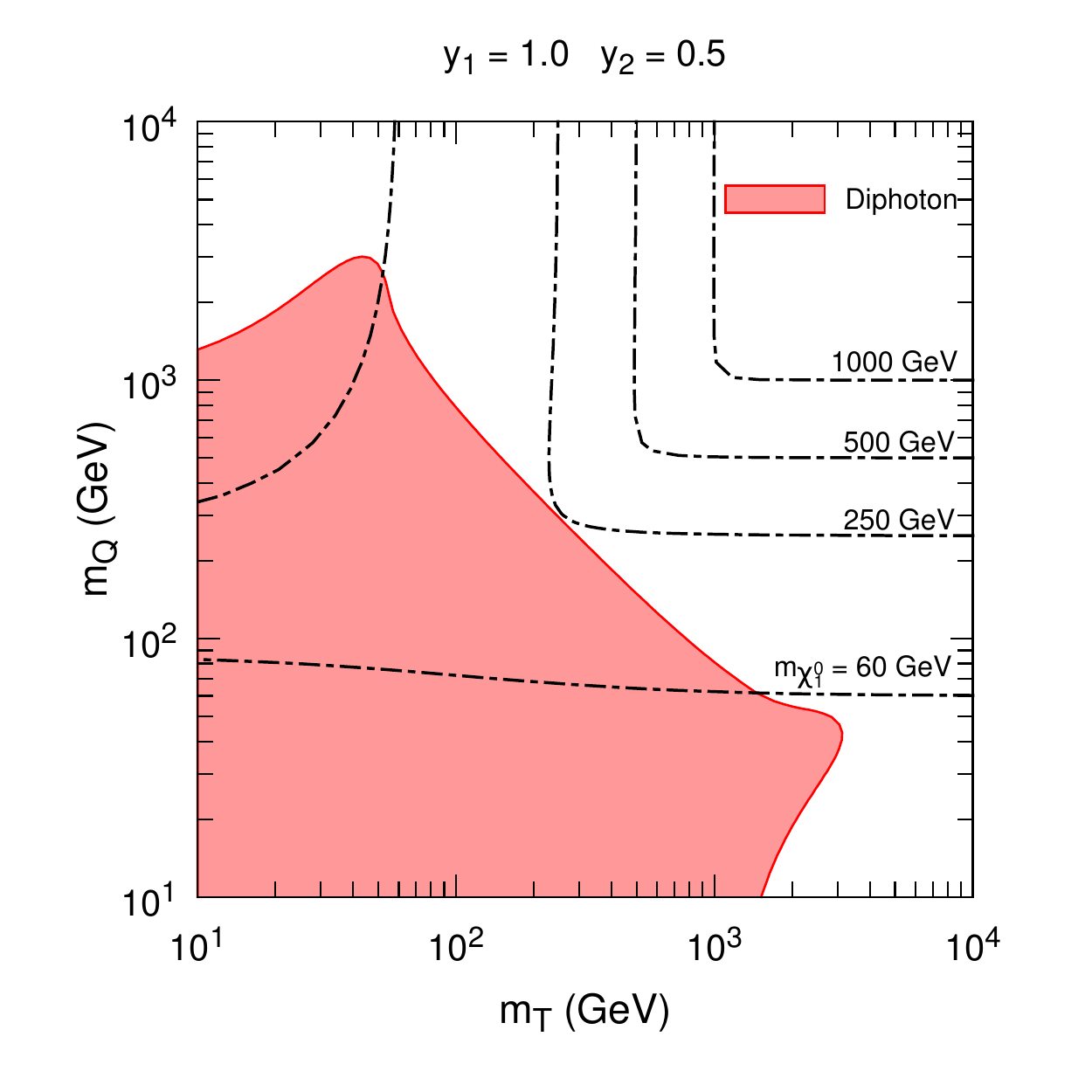}}
\caption{Expected 95\% exclusion regions in the $m_T$-$m_Q$ plane from the CEPC $h\to\gamma\gamma$ measurement with an integrated luminosity of $3~\si{ab^{-1}}$ for the fixed Yukawa couplings of $y_1=y_2=0.5$ (a) and $(y_1,y_2)=(1.0,0.5)$ (b).
Dot-dashed lines denote contours of $m_{\chi_1^0}$.}
\label{diphoton}
\end{figure}

Using the above formula, we calculate the prediction for $\kappa_\gamma$ in the TQDM model and compare it with the CEPC precision.
In Fig.~\ref{diphoton}, the red regions are expected to be excluded by the CEPC $h\to\gamma\gamma$ measurement at 95\% C.L.
We can see that it is possible to explore up to $\chi_1^0\sim$ 200-250 GeV through this measurement.
Compared with the results for the $e^+e^-\to Zh$ measurement shown in Figs.~\ref{eehz:a} and ~\ref{eehz:b}, we find that the $h\to\gamma\gamma$ measurement has less capability to investigate the parameter regions with $m_Q \gg m_T$ or $m_T \gg m_Q$.

\section{Conclusions And Discussions}
\label{sec:conslusion}

In this paper, we investigate the current constraints on the TQDM model from 13~TeV LHC searches, and further study the prospects of searches at future colliders, including the SPPC and the CEPC.
As the dark sector fermions could be directly produced at high energy hadron colliders, we discuss three signal channels, \textit{i.e.}, the $\text{monojet}+\missET$, disappearing track, and $\text{multilepton}+\missET$ channels, at the LHC and the SPPC.

In the $\text{monojet}+\missET$ channel, we find that when $\chi^0_1$ is almost pure triplet (quadruplet), the LHC search has excluded the parameter regions with $m_{\chi^0_1}\lesssim 70~(200)~\si{GeV}$ at 95\% C.L.
Because of the extremely high collision energy, the $\text{monojet}+\missET$ search at the SPPC will be able to explore most of the parameter regions allowed by the observed DM relic density, up to $m_{\chi^0_1}\sim$ 1000-2000 GeV.

If $\chi^\pm_1$ is nearly degenerate in mass with $\chi^0_1$, it would have a moderate lifetime that leads to the disappearing track signal at colliders.
This case can be realized in the regions with $|y_1|=|y_2|$, or  $m_T \gg \max(m_Q, |y_1|v, |y_2|v)$, or $m_Q \gg \max(m_T, |y_1|v, |y_2|v)$.
We find that for a pure triplet (quadruplet) $\chi^0_1$, the parameter regions with $m_{\chi^0_1}\lesssim 410~(472)~\si{GeV}$ are excluded at 95\% C.L. by the disappearing track search at the LHC, while the SPPC would explore up to $m_{\chi^0_1} \sim$ 3200-4500~(3500-5200) GeV.

The $\text{multilepton}+\missET$ channel is suitable to investigate the parameter regions where the mass spectrum is not compressed.
We find that some regions with $m_{\chi^0_1}\lesssim 200~\si{GeV}$ have been excluded by the LHC $\text{multilepton}+\missET$ searches, while the same kind of searches at the SPPC will probe the parameter regions up to $m_{\chi^0_1}\sim 2000~\si{GeV}$
If the mass spectrum is compressed, the discovery capability of this channel is much weaker than that of the $\text{monojet}+\missET$ channel.

On the other hand, the future Higgs factory CEPC will be able to study loop effects of BSM physics through high precision Higgs measurements. We calculate the loop correction to the $e^+e^-\to Zh$ production cross section induced by dark sector fermions in the TQDM model, and find that the related CEPC measurement would explore up to $m_{\chi^0_1}\sim 300~\si{GeV}$ for moderate values of $y_1$ and $y_2$.
We also compute the deviation of the $h\to \gamma\gamma$ partial width induced by $\chi^\pm_{1,2,3}$.
The sensitivity of the $h\to \gamma\gamma$ measurement will be weaker than that of the $e^+e^-\to Zh$ measurement, but it covers some particular regions where the latter loses sensitivity due to the threshold effect.

Although we have treated the Yukawa couplings $y_1$ and $y_2$ as free parameters in the above calculations, large $y_1$ and $y_2$ may cause a problem.
As the Yukawa couplings give negative contributions to the $\beta$ function of the Higgs quartic coupling $\lambda$, sufficient large $y_1$ and $y_2$ may render $\lambda$ negative at a scale much lower than the Planck scale, endangering the stability of the EW vacuum.
In order to evaluate such an effect, we derive the dark sector contributions to the $\beta$ functions of $\lambda$, $g$, $g'$, and $y_t$, the top Yukawa coupling, at one-loop level as
\begin{eqnarray}
\Delta \beta_\lambda &=& \frac{1}{16\pi^2} \left[8\lambda (y_1^2 + y_2^2) - \frac{{28}}{9}~(y_1^4 + y_2^4) - \frac{{40}}{9}y_1^2y_2^2\right],\\
\Delta \beta_g &=& \frac{1}{16\pi^2}\, 8g^3,\quad
\Delta \beta_{g'} = \frac{1}{16\pi^2} \frac{4}{3}{g'}^3,\quad
\Delta \beta_{y_t} = \frac{1}{16\pi^2}\, 2{y_t}(y_1^2 + y_2^2),
\end{eqnarray}
while the $\beta$ functions of $y_1$ and $y_2$ are
\begin{eqnarray}
{\beta _{{y_1}}} &=& \frac{1}{16\pi^2}\; {y_1}\left( {\frac{{19}}{6}y_1^2 + \frac{{10}}{3}y_2^2 - \frac{{69}}{4}{g^2} - \frac{3}{4}{{g'}^2} + 3y_t^2} \right),\\
{\beta _{{y_2}}} &=& \frac{1}{16\pi^2}\; {y_2}\left( {\frac{{19}}{6}y_2^2 + \frac{{10}}{3}y_1^2 - \frac{{69}}{4}{g^2} - \frac{3}{4}{{g'}^2} + 3y_t^2} \right).
\end{eqnarray}
These expressions are derived by hand and crosschecked using \texttt{PyR@TE~2.0.0}~\cite{Lyonnet:2016xiz}.

By solving the renormalization group equations with the initial values of the couplings at the top mass pole~\cite{Buttazzo:2013uya}, we obtain the running values of the couplings at high scales
According to our calculation, if $\sqrt{y^2_1+y^2_2} \lesssim 0.5$, the EW vacuum would be stable up to the Planck scale.
If $0.5 \lesssim \sqrt{y^2_1+y^2_2} \lesssim 0.7$, the vacuum would be metastable.
For an even larger $\sqrt{y^2_1+y^2_2}$, some additional bosonic degrees of freedom would be needed above the TeV scale for ensuring the vacuum stability.
Surprisingly, introducing new Yukawa couplings in the TQDM model do not make the vacuum stability worse than the SM.
The reason is that the large, positive contribution to $\beta_g$ from the triplet and quadruplets increase $g$ at high scales, and hence indirectly lift up $\lambda$.
Nevertheless, $g$ would reach a Landau pole around the Planck scale. Thus, one may expect that there is other new physics below the Planck scale.

\begin{acknowledgments}
This work is supported by the National Natural Science Foundation of China
under Grants No.~11475189 and 11475191, and by the National Key Program for Research and Development (No.~2016YFA0400200).
Z.H.Y. is supported by the Australian Research Council.
\end{acknowledgments}

\bibliographystyle{utphys}
\bibliography{cite_temp}

\end{document}